\documentclass[ALICE,manyauthors]{cernphprep}
\usepackage[comma,square,numbers,sort&compress]{natbib}
\usepackage{hyperref}
\usepackage{lineno}
\usepackage{xspace}
\usepackage{booktabs}
\usepackage{placeins}
\usepackage{enumitem}
\usepackage[T1]{fontenc}
\begin{document}
%

\newcommand{\pp}           {pp\xspace}
\newcommand{\ppbar}        {\mbox{$\mathrm {p\overline{p}}$}\xspace}
\newcommand{\XeXe}         {\mbox{Xe--Xe}\xspace}
\newcommand{\PbPb}         {\mbox{Pb--Pb}\xspace}
\newcommand{\pA}           {\mbox{pA}\xspace}
\newcommand{\pPb}          {\mbox{p--Pb}\xspace}
\newcommand{\AuAu}         {\mbox{Au--Au}\xspace}
\newcommand{\dAu}          {\mbox{d--Au}\xspace}

\newcommand{\s}            {\ensuremath{\sqrt{s}}\xspace}
\newcommand{\snn}          {\ensuremath{\sqrt{s_{\mathrm{NN}}}}\xspace}
\newcommand{\pt}           {\ensuremath{p_{\rm T}}\xspace}
\newcommand{\meanpt}       {$\langle p_{\mathrm{T}}\rangle$\xspace}
\newcommand{\ycms}         {\ensuremath{y_{\rm CMS}}\xspace}
\newcommand{\ylab}         {\ensuremath{y_{\rm lab}}\xspace}
\newcommand{\etarange}[1]  {\mbox{$\left | \eta \right |~<~#1$}}
\newcommand{\yrange}[1]    {\mbox{$\left | y \right |~<~#1$}}
\newcommand{\dndy}         {\ensuremath{\mathrm{d}N_\mathrm{ch}/\mathrm{d}y}\xspace}
\newcommand{\dndeta}       {\ensuremath{\mathrm{d}N_\mathrm{ch}/\mathrm{d}\eta}\xspace}
\newcommand{\avdndeta}     {\ensuremath{\langle\dndeta\rangle}\xspace}
\newcommand{\dNdy}         {\ensuremath{\mathrm{d}N_\mathrm{ch}/\mathrm{d}y}\xspace}
\newcommand{\Npart}        {\ensuremath{N_\mathrm{part}}\xspace}
\newcommand{\Ncoll}        {\ensuremath{N_\mathrm{coll}}\xspace}
\newcommand{\dEdx}         {\ensuremath{\textrm{d}E/\textrm{d}x}\xspace}
\newcommand{\RpPb}         {\ensuremath{R_{\rm pPb}}\xspace}

\newcommand{\nineH}        {$\sqrt{s}~=~0.9$~Te\kern-.1emV\xspace}
\newcommand{\seven}        {$\sqrt{s}~=~7$~Te\kern-.1emV\xspace}
\newcommand{\twoH}         {$\sqrt{s}~=~0.2$~Te\kern-.1emV\xspace}
\newcommand{\twosevensix}  {$\sqrt{s}~=~2.76$~Te\kern-.1emV\xspace}
\newcommand{\five}         {$\sqrt{s}~=~5.02$~Te\kern-.1emV\xspace}
\newcommand{\twosevensixnn}{$\sqrt{s_{\mathrm{NN}}}~=~2.76$~Te\kern-.1emV\xspace}
\newcommand{\fivenn}       {$\sqrt{s_{\mathrm{NN}}}~=~5.02$~Te\kern-.1emV\xspace}
\newcommand{\LT}           {L{\'e}vy-Tsallis\xspace}
\newcommand{\GeVc}         {Ge\kern-.1emV/$c$\xspace}
\newcommand{\MeVc}         {Me\kern-.1emV/$c$\xspace}
\newcommand{\TeV}          {Te\kern-.1emV\xspace}
\newcommand{\GeV}          {Ge\kern-.1emV\xspace}
\newcommand{\MeV}          {Me\kern-.1emV\xspace}
\newcommand{\GeVmass}      {Ge\kern-.2emV/$c^2$\xspace}
\newcommand{\MeVmass}      {Me\kern-.2emV/$c^2$\xspace}
\newcommand{\lumi}         {\ensuremath{\mathcal{L}}\xspace}

\newcommand{\ITS}          {\rm{ITS}\xspace}
\newcommand{\TOF}          {\rm{TOF}\xspace}
\newcommand{\ZDC}          {\rm{ZDC}\xspace}
\newcommand{\ZDCs}         {\rm{ZDCs}\xspace}
\newcommand{\ZNA}          {\rm{ZNA}\xspace}
\newcommand{\ZNC}          {\rm{ZNC}\xspace}
\newcommand{\SPD}          {\rm{SPD}\xspace}
\newcommand{\SDD}          {\rm{SDD}\xspace}
\newcommand{\SSD}          {\rm{SSD}\xspace}
\newcommand{\TPC}          {\rm{TPC}\xspace}
\newcommand{\TRD}          {\rm{TRD}\xspace}
\newcommand{\VZERO}        {\rm{V0}\xspace}
\newcommand{\VZEROA}       {\rm{V0A}\xspace}
\newcommand{\VZEROC}       {\rm{V0C}\xspace}
\newcommand{\Vdecay} 	   {\ensuremath{V^{0}}\xspace}

\newcommand{\ee}           {\ensuremath{e^{+}e^{-}}} 
\newcommand{\pip}          {\ensuremath{\pi^{+}}\xspace}
\newcommand{\pim}          {\ensuremath{\pi^{-}}\xspace}
\newcommand{\kap}          {\ensuremath{\rm{K}^{+}}\xspace}
\newcommand{\kam}          {\ensuremath{\rm{K}^{-}}\xspace}
\newcommand{\pbar}         {\ensuremath{\rm\overline{p}}\xspace}
\newcommand{\kzero}        {\ensuremath{{\rm K}^{0}_{\rm{S}}}\xspace}
\newcommand{\lmb}          {\ensuremath{\Lambda}\xspace}
\newcommand{\almb}         {\ensuremath{\overline{\Lambda}}\xspace}
\newcommand{\Om}           {\ensuremath{\Omega^-}\xspace}
\newcommand{\Mo}           {\ensuremath{\overline{\Omega}^+}\xspace}
\newcommand{\X}            {\ensuremath{\Xi^-}\xspace}
\newcommand{\Ix}           {\ensuremath{\overline{\Xi}^+}\xspace}
\newcommand{\Xis}          {\ensuremath{\Xi^{\pm}}\xspace}
\newcommand{\Oms}          {\ensuremath{\Omega^{\pm}}\xspace}
\newcommand{\degree}       {\ensuremath{^{\rm o}}\xspace}

\newcommand{\slfrac}[2]{\left.#1\right/#2}
\newcommand{\Ppsi}{$\rm{\psi'}$}
\newcommand{\Jpsi}{$\rm{J/\psi}$}
\newcommand{\Noon}{\textbf{n$\mathbf{_O^O}$n}}
\newcommand{\mumu}{$\mu^{+} \mu^{-}$}
\newcommand{\elel}{$e^{+} e^{-}$}
\newcommand{\pbp}{$\rm{p} \overline{p}$}
\newcommand{\eepp}{$e^{+} e^{-} \pi^{+} \pi^{-}$}
\newcommand{\mmpp}{$\mu^{+} \mu^{-} \pi^{+} \pi^{-}$}
\newcommand{\llpp}{$l^{+} l^{-} \pi^{+} \pi^{-}$}
\newcommand{\lplp}{$l^{+} l^{-}$}

\begin{titlepage}
\PHyear{2021}       
\PHnumber{002}      
\PHdate{5 January}  

\title{Coherent \Jpsi~and \Ppsi~photoproduction at midrapidity in ultra-peripheral \PbPb~collisions at \fivenn}
\ShortTitle{Coherent \Jpsi~and \Ppsi~photoproduction at midrapidity}   

\Collaboration{ALICE Collaboration\thanks{See Appendix~\ref{app:collab} for the list of collaboration members}}
\ShortAuthor{ALICE Collaboration} 

\begin{abstract}
The coherent photoproduction of \Jpsi\ and \Ppsi\ mesons was measured in ultra-peripheral \PbPb\ collisions at a center-of-mass energy \fivenn~with the ALICE detector. Charmonia are detected in the central rapidity region for events where the hadronic interactions are strongly suppressed. 
The \Jpsi\ is reconstructed using the dilepton (\lplp) and proton-antiproton decay channels, while
for the \Ppsi\, the dilepton and the \llpp\ decay channels are studied.
The analysis is based on an event sample corresponding to an integrated luminosity of about 233 ${\rm \mu b}^{-1}$.  The results are compared with theoretical models for coherent \Jpsi\ and \Ppsi\ photoproduction. The coherent cross section
is found to be in a good agreement with models incorporating moderate nuclear gluon shadowing of about 0.64 at a Bjorken-$x$ of around $6\times 10^{-4}$, such as the EPS09 parametrization, however none of the models is able to fully describe the rapidity dependence of the coherent \Jpsi\ cross section including ALICE measurements at forward rapidity. The ratio of \Ppsi\ to \Jpsi\ coherent photoproduction cross sections was also measured and found to be consistent with the one for photoproduction off protons.

\end{abstract}
\end{titlepage}

\setcounter{page}{2} 


\section{Introduction} 

Photonuclear reactions
can be studied in ultra-peripheral collisions (UPCs) of heavy ions where the two nuclei pass by with an impact parameter larger than the sum of their radii. Hadronic interactions are suppressed and the dominant electromagnetic interactions are mediated by photons of small virtualities. The intensity of the photon flux is growing with the squared nuclear charge  of the colliding ion resulting in large cross sections for the photoproduction of vector mesons in heavy-ion collisions. The photoproduction process has a clear experimental signature: the decay products of vector mesons are the only signal in an otherwise empty detector.

The physics of vector meson photoproduction is described in~\cite{Bertulani:2005ru, Baltz:2007kq,Contreras:2015dqa,Klein:2019qfb}.  Photoproduction of vector mesons in ion collisions can either be coherent, i.e. the photon interacts consistently with all nucleons in a nucleus, or incoherent, i.e. the photon interacts with a single nucleon. Experimentally, one can distinguish between these two production types through the typical transverse momentum of the produced vector mesons, which is inversely proportional to the transverse size of the target. While the coherent photoproduction is characterized by the production of mesons with low transverse momentum ($\langle \pt \rangle \sim$~60~MeV/$c$), the incoherent is dominated by mesons with higher values ($\langle \pt \rangle \sim$~500~MeV/$c$). In the first case, the nuclei usually do not dissociate, but the electromagnetic fields of ultrarelativistic heavy nuclei are strong enough to develop other independent soft electromagnetic interactions accompanying  the coherent photoproduction process and resulting in the excitation of one or both of the nuclei. In the second case, the nucleus breaks up and usually emits neutrons close to the beam rapidities which can be measured in zero-degree calorimeters (ZDC) placed at long distances on both sides of the detector\cite{ALICE:2012aa}. The incoherent photoproduction can also be accompanied by the excitation and dissociation of the target nucleon resulting in even higher transverse momenta of the produced vector mesons~\cite{Guzey:2018tlk}.

Coherent heavy vector meson photoproduction is of particular interest because of its connection with the gluon distribution functions (PDFs)  in protons and nuclei~\cite{Ryskin:1992ui}. At low Bjorken-$x$ values, nuclear parton distribution functions are significantly suppressed in the nucleus with respect to free proton PDFs, a phenomenon known as parton shadowing~\cite{Armesto:2006ph}. Shadowing effects are usually attributed to multiple scattering and addressed in various phenomenological approaches based on elastic Glauber-like rescatterings of hadronic components of the photon, Glauber-Gribov inelastic rescatterings, and high-density QCD~\cite{Frankfurt:2011cs,Bendova:2020hbb,PhysRevC.90.015203,PhysRevC.83.065202,Luszczak:2019vdc,Cepila:2017nef}. Besides, different parameterizations of nuclear partonic distributions based on fits to existing data are available~\cite{Eskola:2009uj,Eskola:2016oht,Kovarik:2015cma,AbdulKhalek:2020yuc}, however these parameterizations are affected by large uncertainties at low Bjorken-$x$ values due to the limited kinematic coverage of the available data samples.

Heavy vector meson photoproduction measurements provide a powerful tool to study poorly known gluon shadowing effects at low $x$.
The scale of the four-momentum transfer of the interaction is related to the mass $m_V$ of the vector meson as $Q^{2}~\sim~m^{2}_{V}/4$ corresponding to the perturbative regime in the case of heavy charmonium states. The rapidity of the coherently produced $\rm c \bar c$ states is related to the Bjorken-$x$ of the gluons as $x~=~\left(m_V/\sqrt{s_{NN}}\right)\exp\left(\pm~y\right)$, where the sign of the exponent reflects that each of the incoming lead nuclei may act as the photon source. The gluon shadowing factor $R_g(x,Q^2)$, i.e. the ratio of the nuclear gluon density distribution to the gluon distribution in the proton, can be evaluated via the measurement of the nuclear suppression factor defined as the square root of the ratio of the coherent vector meson photoproduction cross section on nuclei to the photoproduction cross section in the impulse approximation that is based on the exclusive photoproduction measurements with the proton target~\cite{Guzey:2013xba,Contreras:2016pkc}. The square root in this definition is motivated by the fact that the coherent photoproduction cross section is expected to scale as the square of the gluon density in leading order pQCD.

The extraction of the nuclear suppression factor in UPC measurements is complicated by the fact that the measured vector meson cross section in UPCs is expressed as a sum of two contributions since either of the colliding ions can serve as a photon source. At forward rapidities one contribution corresponds to higher photon--nucleus energies while the other to lower energies resulting in ambiguities in the extraction of the nuclear suppression factor. The midrapidity region has the advantage that both contributions are the same and the suppression factor can be extracted unambiguously in this case. Since the contribution of the high and low energies as the rapidity changes  varies across models, the measurement of the rapidity dependence of the cross section for coherent charmonium production may provide a new tool to constrain the evolution of the parton distribution in the different models.

Photoproduction cross section measurements for  different quarkonium species provide an opportunity to probe gluon shadowing effects at different $Q^2$ scales. On the other hand, the comparison of excited and ground states of charmonia can shed light on the modelling of the charmonium wave functions and help to disentangle perturbative from non-perturbative effects in the model calculations~\cite{Cepila:2019skb, Krelina:2020bxt}.

Charmonium photoproduction in \PbPb\ UPCs was previously studied by the ALICE Collaboration at \twosevensixnn~\cite{Abelev:2012ba, Abbas:2013oua,Adam:2015sia}. The coherent \Jpsi\ photoproduction cross section was measured both at midrapidity $|y|<0.9$ and at forward rapidity $-3.6 < y < -2.6$. In addition, the CMS Collaboration studied the coherent \Jpsi\ photoproduction accompanied by neutron emission at semi-forward rapidity $1.8 < |y| < 2.3$ at \twosevensixnn~\cite{Khachatryan:2016qhq}. The results were compared with various models and the best description was found amongst those introducing moderate gluon shadowing in the nucleus. 
The ALICE measurements were used in Ref.~\cite{Guzey:2013xba} to extract the nuclear gluon shadowing factor $R_g$ yielding $R_g(x \sim 10^{-3}) = 0.61^{+0.05}_{-0.04}$ and $R_g(x \sim 10^{-2}) = 0.74^{+0.11}_{-0.12}$ at the scale of the charm quark mass. The ALICE measurement of \Ppsi\ photoproduction at midrapidity also supports the moderate-shadowing scenario~\cite{Adam:2015sia}. A complementary rapidity-differential measurement of the coherent \Jpsi\ and \Ppsi\ photoproduction at forward rapidity in \PbPb\ UPCs at \fivenn by the ALICE Collaboration further underlines the importance of gluon shadowing effects~\cite{Acharya:2019vlb}.
The gluon shadowing factor $R_g(x \sim 10^{-2}) \sim 0.8$ was obtained under assumption that the contribution from high photon-nucleus energies, i.e. low Bjorken $x\sim 10^{-5}$, can be neglected in the measured cross sections.

In this publication, we present the first measurement of the coherent \Jpsi\ and \Ppsi\ photoproduction cross sections at the midrapidity range $|y|<0.8$ in the Pb--Pb UPCs at \fivenn, recorded by ALICE in 2018.  The \Jpsi\ photoproduction cross section in this measurement is sensitive to $x\in(0.3,1.4)\times10^{-3}$, a factor 2 smaller than in the previous midrapidity measurement at \twosevensixnn~\cite{Abbas:2013oua}. This data sample is approximately 10 times larger than Pb--Pb sample at \twosevensixnn used for the ALICE results reported in Refs.~\cite{Abbas:2013oua,Adam:2015sia}. The larger data sample allows for a measurement of the \Jpsi\ cross section in three rapidity intervals ($|y|<0.15$, $0.15<|y|<0.35$, $0.35<|y|<0.8$) extending the previous rapidity-differential cross section measurement in the forward range  at \fivenn~\cite{Acharya:2019vlb}. \Jpsi\ decays to \mumu, \elel\ and \pbp\ and \Ppsi\ decays to \mmpp, \eepp\ and \lplp\ are investigated. The coherent \Jpsi\ production in the \pbp\ channel in UPCs is measured for the first time. The ratio of the \Ppsi\ and \Jpsi\ cross sections is also measured and compared with earlier ALICE measurements~\cite{Adam:2015sia,Acharya:2019vlb}. The measured cross sections are compared to models assuming no gluon shadowing as well as to predictions that employ moderate gluon shadowing. Shadowing models are based on a parametrization of previously available data, the leading twist approximation and several variations of the color dipole approach.

\section{Detector description}

The ALICE detector and its performance are described in~\cite{Aamodt:2008zz, Abelev:2014ffa}. The main components of the ALICE detector are a central barrel placed in a large solenoid magnet ($B=0.5$~T), covering the central pseudorapidity region, and a muon spectrometer at forward rapidity, covering the range $-4.0 <\eta<-2.5$. Three central barrel detectors, the Inner Tracking System (ITS), the Time Projection Chamber (TPC), and the Time-of-Flight detector (TOF), are used in this analysis. 

The ITS is made of six silicon layers and is used for particle tracking and interaction vertex reconstruction~\cite{Aamodt:2010aa}. The Silicon Pixel Detector (SPD) makes up the two innermost layers of the ITS with about $10^7$ pixels covering the pseudorapidity intervals $|\eta| < 2$ and $|\eta| < 1.4$ for the inner (radius 3.9 cm) and outer (radius 7.6 cm) layers, respectively. The SPD is read out by 400 (800) chips in the inner (outer) layer with each of the readout chips also providing a trigger signal if at least one of its pixels is fired. When projected into the transverse plane, the chips are arranged in 20 (40) azimuthal regions in the inner (outer) layer allowing for a topological selection of events at the trigger level.

The TPC is used for tracking and for particle identification~\cite{Alme_2010}. A 100 kV central electrode separates the two drift volumes, providing an electric field for electron drift. The two end-plates, at $|z| = 250$ cm, are instrumented with Multi-Wire-Proportional-Chambers (MWPCs) with 560,000 readout pads, allowing high precision track measurements in the transverse plane. The $z$ coordinate is given by the time of drift in the TPC electric field. The TPC acceptance covers the pseudorapidity region $|\eta| <0.9$. Ionization measurements of individual track clusters are used for particle identification.

The TOF detector is a large cylindrical barrel of multigap resistive plate chambers with about 150,000 readout channels surrounding the TPC and providing very high precision timing measurement~\cite{Akindinov:2013tea}. The TOF pseudorapidity coverage is $|\eta| <0.8$. In combination with the tracking system, the TOF detector is used for charged particle identification up to a  momentum of about 2.5 GeV$/c$ for pions and kaons  and up to  4 GeV$/c$ for protons. The TOF readout channels are grouped into 1608 trigger channels (maxipads) arranged into 18 azimuthal regions and provide topological-trigger decisions.

The measurement also makes use of the three forward detectors. The V0 counters consist of two arrays of 32 scintillator tiles each, covering the interval $2.8\rm<\eta<5.1$ (V0A) and $-3.7\rm<\eta<-1.7$ (V0C) and positioned respectively at $z=340$~cm and $z=-90$~cm from the interaction point~\cite{Abbas:2013taa}.  
The ALICE Diffractive (AD) detector consists of two arrays of 8 scintillator tiles each arranged in two layers, covering the range $4.9\rm<\eta<6.3$ (ADA) and $-7.0\rm<\eta<-4.8$ (ADC) and positioned at $z=17$~m and $z=-19.5$~m from the interaction point, respectively~\cite{Akiba:2016ofq}. Both V0 and AD can be used to veto hadronic interactions at the trigger level.  

Finally, two zero-degree calorimeters ZNA and ZNC, located at $\pm112.5$~m from the interaction point, are used for the measurement of neutrons at beam rapidity. They have good efficiency ($\approx0.94$) to detect neutrons with $|\eta|>8.8$ and have a relative energy resolution of about 20\% for single neutrons, which allows for a clear separation of events with either zero or a few neutrons at beam rapidity~\cite{ALICE:2012aa}.

\section{Data analysis}
\subsection{Event selection}
The data analysis in this paper is based on the event sample recorded during the \PbPb~at \fivenn data taking period in 2018. A dedicated central barrel UPC trigger consists of topological trigger formed by at least two and up to six TOF maxipads with at least one pair of maxipads having an opening angle in azimuth larger than 150 degrees and a topological trigger formed by at least four triggered SPD chips. The triggered SPD chips are required to form two pairs, each pair with two chips in different SPD layers falling in compatible azimuthal regions. The two pairs of chips are required to have an opening angle in azimuth larger than 153 degrees. It is further vetoed by any activity within the time windows for nominal beam–beam interactions on the V0 and AD detectors on both sides of the interaction point.

The used data sample corresponds to an integrated luminosity of $233\, \mu{\rm b}^{-1}$, derived from the counts of two independent reference triggers, one was based on multiplicity selection in the V0 detector and another one based on neutron detection in the ZDC. The reference trigger cross sections were determined from van der Meer scans; this procedure has an uncertainty of 2.2\%~\cite{ALICE-PUBLIC-2021-001}.The determination of the live-time of the UPC trigger has an additional uncertainty of 1.5\%. The total  relative systematic uncertainty of the integrated luminosity is thus 2.7\%. Alternatively the reference cross section can be calibrated using Glauber calculation~\cite{Acharya:2019vlb}. The difference between the two methods is smaller than the quoted systematic uncertainty. Alternatively the reference cross section can be calibrated using Glauber calculation~\cite{Acharya:2019vlb}. The difference between the two methods is smaller than the quoted systematic uncertainty.

Additional offline vetoes are applied on the AD and V0 detector signals to ensure the exclusive production of the charmonia. The offline selection in these detectors is more precise than vetoes at the trigger level, because it relies on larger time windows than the trigger electronics and on a more refined algorithm to quantify the signal.

Online and offline V0 and AD veto requirements may result in significant inefficiencies (denoted as veto inefficiencies) in selecting signal events with exclusive charmonium production due to additional activity induced by hadronic or electromagnetic pile–up processes from independent Pb--Pb collisions accompanying the coherent charmonium photoproduction. The probability of hadronic pile–up in the collected sample does not exceed 0.2\%, however there is a significant pile–up contribution from the electromagnetic electron-pair production process. The veto inefficiency induced by these pile–up effects in the V0 and AD detectors is estimated using events selected with an unbiased trigger based only on the timing of bunches crossing the interaction region. The average veto efficiency $\epsilon_{\rm veto}^{\rm pileup}=0.920 \pm 0.002$ is applied to raw charmonium yields to account for hadronic and electromagnetic pile-up processes. In the forward rapidity analysis a higher veto efficiency was obtained due to missing veto on V0C detector.

Signal events with exclusive charmonium production, accompanied by electromagnetic nuclear dissociation (EMD), can be rejected if, in addition to the forward neutrons, other particles, produced at large rapidities, leave a signal either in the AD or the V0 detectors. These extra particles may come from multifragmentation or pion production processes, and the corresponding cross sections are expected to be large~\cite{Pshenichnov:2011zz}. The amount of good events with neutrons, which are lost due to  AD and V0 vetoes, is estimated using control triggers without veto on AD and/or VZERO detectors. The fraction of losses for this category of events (EMD) amounts to $26\% \pm 4\%$ for events with a signal either in ZNA or ZNC and reach $43\% \pm 5\%$ for events with a signal in both ZNA and ZNC. The average event loss is computed using fractions of events with and without neutrons on either side. The average veto efficiency correction $\epsilon_{\rm veto}^{\rm EMD} = 0.92 \pm 0.02$ is applied to raw charmonium yields to account for the EMD process.

The selected events are required to have a reconstructed primary vertex determined using at least two reconstructed tracks and having a longitudinal position within 15 cm of either side of the nominal interaction point.
The analysis is aimed at the reconstruction of \Jpsi\ decaying to \mumu, \elel, \pbp\, and of \Ppsi\ decaying to \lplp and \Jpsi $\pi^+\pi^-$ followed by \Jpsi $\rightarrow l^+ l^-$. Therefore, events with two or four tracks in the central barrel are required. 

Two types of tracks are considered in the analysis: global tracks and ITS standalone tracks.
Global tracks are reconstructed using combined tracking in ITS and TPC detectors. Tracks are required to cross at least 70 (out of 159) TPC pad-rows and to have a cluster on each of the two layers of the SPD. Each track must have a distance of closest approach to the primary vertex of less than 2 cm in the direction of $z$-axis. ITS standalone tracks are reconstructed using ITS clusters not attached to any global track,  requiring at least four clusters in the ITS, out of which two must be in the SPD.

The two-body decays are selected by looking for events with exactly two global tracks with opposite electric charge (unlike-sign). The probability to find extra global tracks not passing the standard track selection criteria or being reconstructed only in ITS is found to be negligible. The four-body decays of \Ppsi\ are selected by looking for exactly four tracks with at least two being global tracks. The kinematics of the $\psi' \rightarrow l^+l^-\pi^+\pi^-$ decay is such that pions and leptons are well separated: leptons have high $\pt \approx 1$\,GeV/$c$ while pions are much softer with $\pt \approx 0.3$\,GeV/$c$. This feature is used to identify the pion pair. Tracks are sorted according to their \pt and the two with lowest \pt are assumed to be pions, while the other two are assumed to be leptons. The tagged pions and lepton pairs are required to consist of opposite-sign tracks.

To separate the \Jpsi $\rightarrow$ \mumu, \elel\ and \pbp\ decays, the particle identification (PID) capabilities of the TPC and TOF detectors are used. 
The momenta of the tracks from \Jpsi\ decays are $p \in (1.0, 2.0)$~GeV$/c$ for the \mumu\ and \elel\ channels and $ p \in (0.75, 1.75)$~GeV$/c$ for the \pbp\ channel. The PID resolution of the TPC allows for complete separation of electrons and muons in the momentum range mentioned above. Since the specific ionization energy loss (${\rm d}E/{\rm d}x$) of electron and proton become equal at momenta around 1 GeV/$c$, the TPC PID is not applicable for the identification of protons from coherently produced \Jpsi. However, the PID capabilities of the TOF detector allow for the separation of protons from other particle species in the momentum range relevant for this analysis. For the \Jpsi $\rightarrow$ \pbp\ channel, at least one track is required to have valid TOF PID information. If no TOF PID is available for the second track, TPC PID is used. The ${\rm d}E/{\rm d}x$ in TPC or the Lorentz Beta factor ($\beta = v/c$) of each reconstructed track in TOF is measured in units of the standard deviation ($\sigma$) with respect to expected values for $\mu,e,\rm p$ at the given measured momentum. The track pair is accepted if $n^{2}_{\sigma +} + n^{2}_{\sigma - } < 16$.  

The charmonium photoproduction may be accompanied by pile-up from electromagnetic electron-pair production or by noise in the SPD resulting in extra fired SPD trigger chips satisfying the SPD trigger selection topology. In order to exclude contamination of events  not triggered by the charmonium decay products, the fired SPD trigger chips are required to match SPD clusters corresponding to the selected tracks. It is found that 11\% (7\%) of the events with a \Jpsi\ candidate decaying into dimuons (di-electrons) with 4 SPD clusters cannot be matched to the fired trigger chips. The matching requirement has a much stronger effect for the 4-track decay channels of \Ppsi\ removing 40\%  and 22\% of the candidates in the \Ppsi $\rightarrow$ \mmpp\ and \Ppsi $\rightarrow$ \eepp channel, respectively. 

\subsection{Acceptance and efficiency correction}
The product of acceptance and efficiency of the \Jpsi\ and \Ppsi\ reconstruction ($\epsilon$) is evaluated using a  large Monte Carlo (MC) sample of coherent and incoherent \Jpsi~and \Ppsi~events generated by STARlight 2.2.0~\cite{Klein:2016yzr} with decay particles tracked in a model of the experimental apparatus implemented in GEANT 3.21~\cite{Brun:1082634}. The model includes a realistic description of the detector status during data taking and its variation with time. 

For this analysis, the primary \Jpsi\ and \Ppsi\ vector mesons produced in UPCs are considered to be transversely polarized. This is consistent with expectations from helicity conservation in photo production and consistent with H1 and ZEUS measurements~\cite{Alexa:2013xxa,Adloff:2002re,Chekanov:2002xi}.
As observed in previous experiments, both \Jpsi\ and the two pions from \Ppsi\ decay are in the S-wave state resulting into the full transfer of the \Ppsi\ polarisation to the \Jpsi ~\cite{PhysRevD.62.032002}. The expected polarization states of primary \Jpsi\ and \Ppsi\ as well as of secondary \Jpsi\ from \Ppsi\ decays are properly taken into account in the MC simulations used in this analysis. These MC simulations are also used in the evaluation of the feed-down contribution to the two-body decay channels from the $\psi' \rightarrow \rm J/\psi + \pi^{+} \pi^{-}$ and $\rm \psi' \rightarrow \rm J/\psi + \pi^{0} \pi^{0}$ decays and for modeling the signal shape and different background contributions. 

The efficiency of the SPD trigger chips is measured with a data-driven approach using a minimum bias trigger. Tracks selected without requiring hits in both SPD layers are matched to the trigger chips they cross. The obtained efficiency maps are introduced on an event-by-event basis to the MC simulations. The single chip efficiency is about 92\%. As the trigger requires 4 chips, the overall effect corresponds to an efficiency of about 0.72 $\pm$ 0.01.

The TOF trigger efficiency is also estimated with a data-driven approach and is taken into account in the MC simulations. The average coverage of the active TOF trigger channels is approximately 90\%. The average trigger efficiency of the active channels is defined as the probability to find signals in maxipads crossed by extrapolated TPC tracks from minimum bias events and is found to be 97--98\%, depending on track arrival times. More than 93\% of the active channels are almost 100\% efficient. The low efficiency in some channels is caused by timing alignment issues and partially disconnected or broken equipment.

\FloatBarrier
\subsection{Signal extraction}
The extraction of coherent \Jpsi\ and \Ppsi\ yields in all decay channels is performed in the rapidity interval $|y|<0.8$. In addition, the \Jpsi\ measurements in the dielectron and dimuon channels are performed in three rapidity intervals: $|y| < 0.15$, $0.15 < |y| < 0.35$, and $0.35 < |y| < 0.8$ where the $y$ ranges were chosen to have approximately the same number of candidates per range. An enriched sample of coherent \Jpsi\ and \Ppsi\ candidates is obtained by selecting the reconstructed candidates with transverse momentum $\pt< 0.2$~GeV$/c$. 

\begin{figure}
\centering
\subfigure[\Jpsi $\rightarrow$ \mumu]
{\label{fig:MassPtJpsi_a}\includegraphics[width=0.49\textwidth]{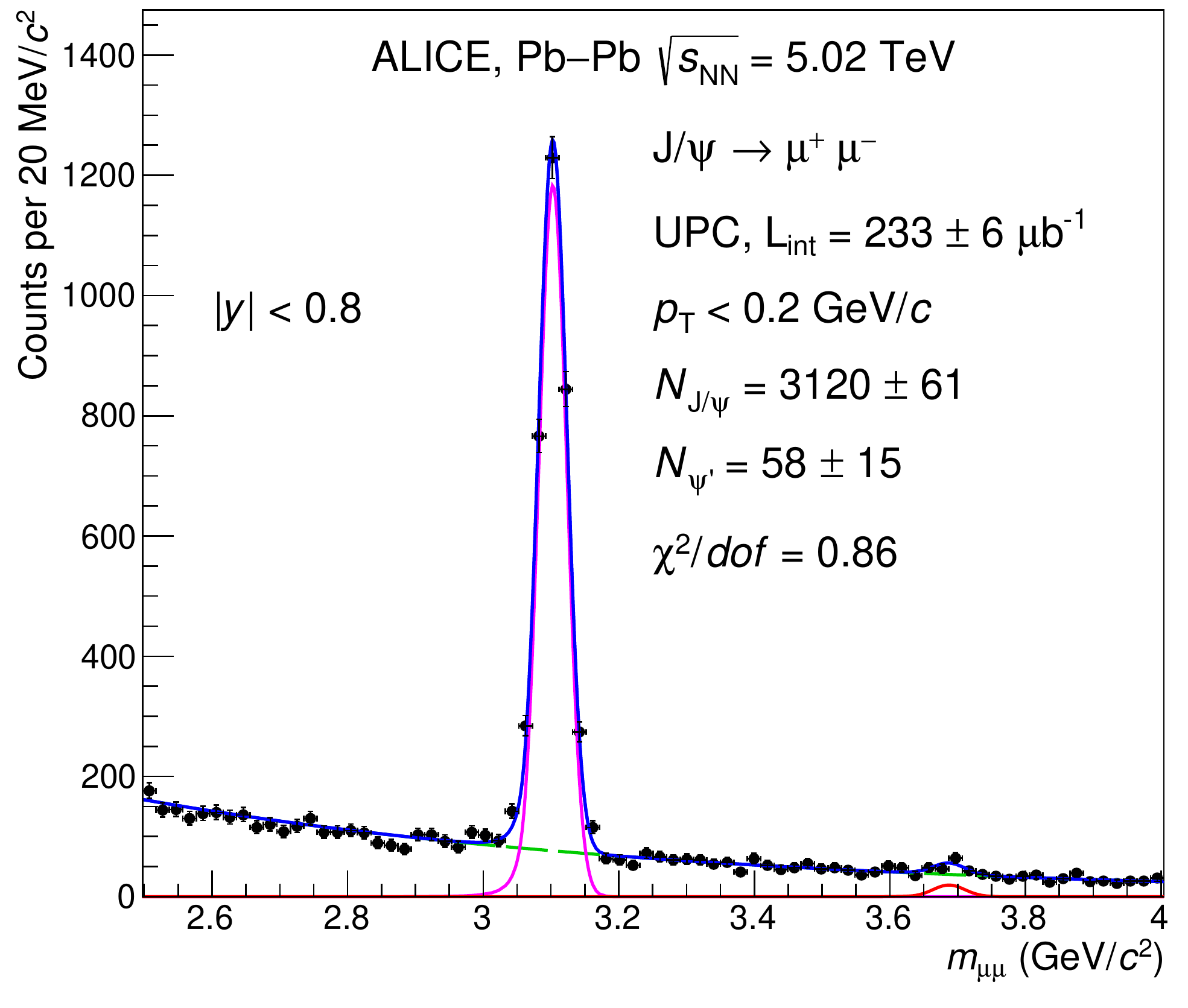}
\includegraphics[width=0.49\textwidth]{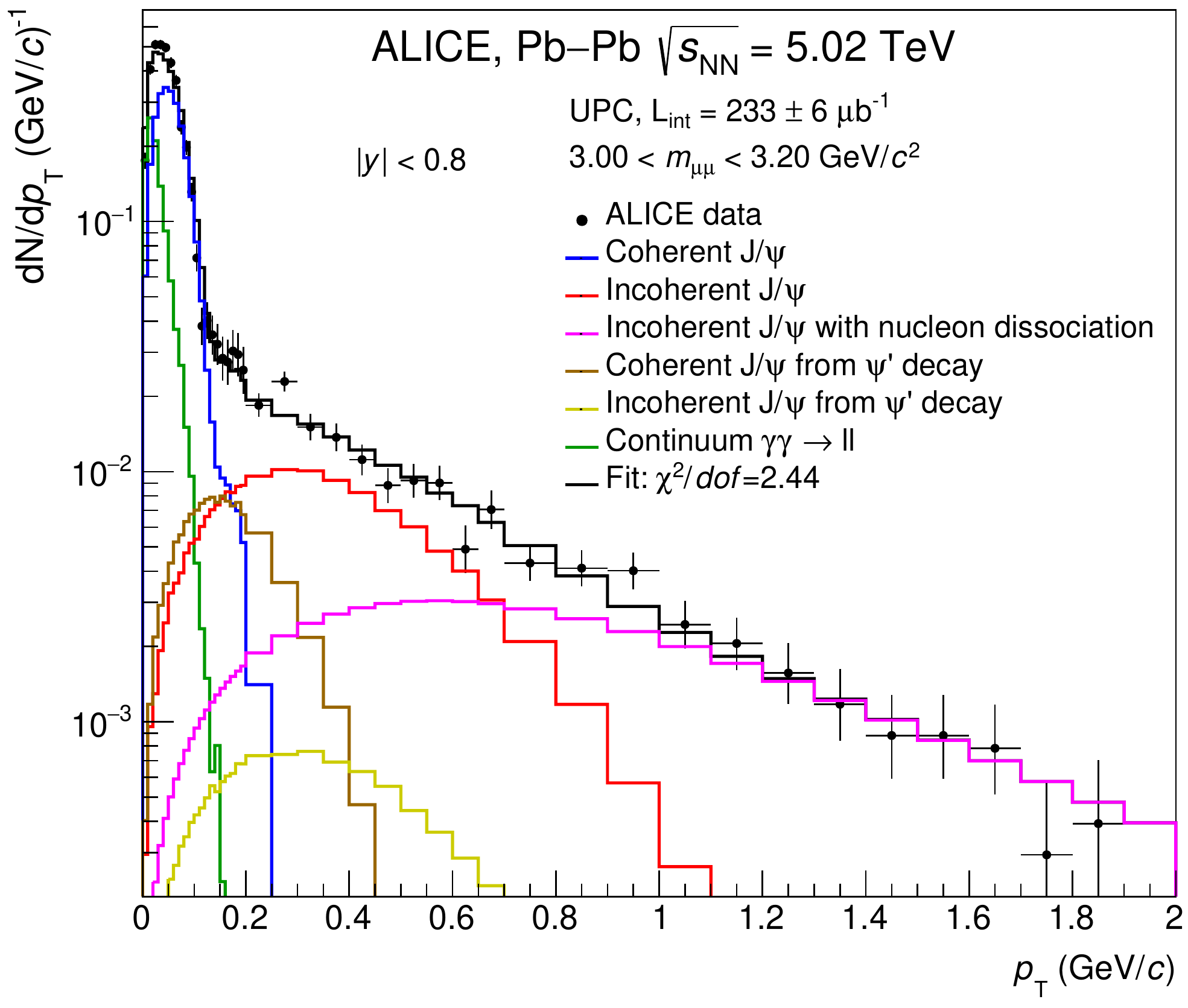}}
\subfigure[\Jpsi $\rightarrow$ \elel]
{\label{fig:MassPtJpsi_b}\includegraphics[width=0.49\textwidth]{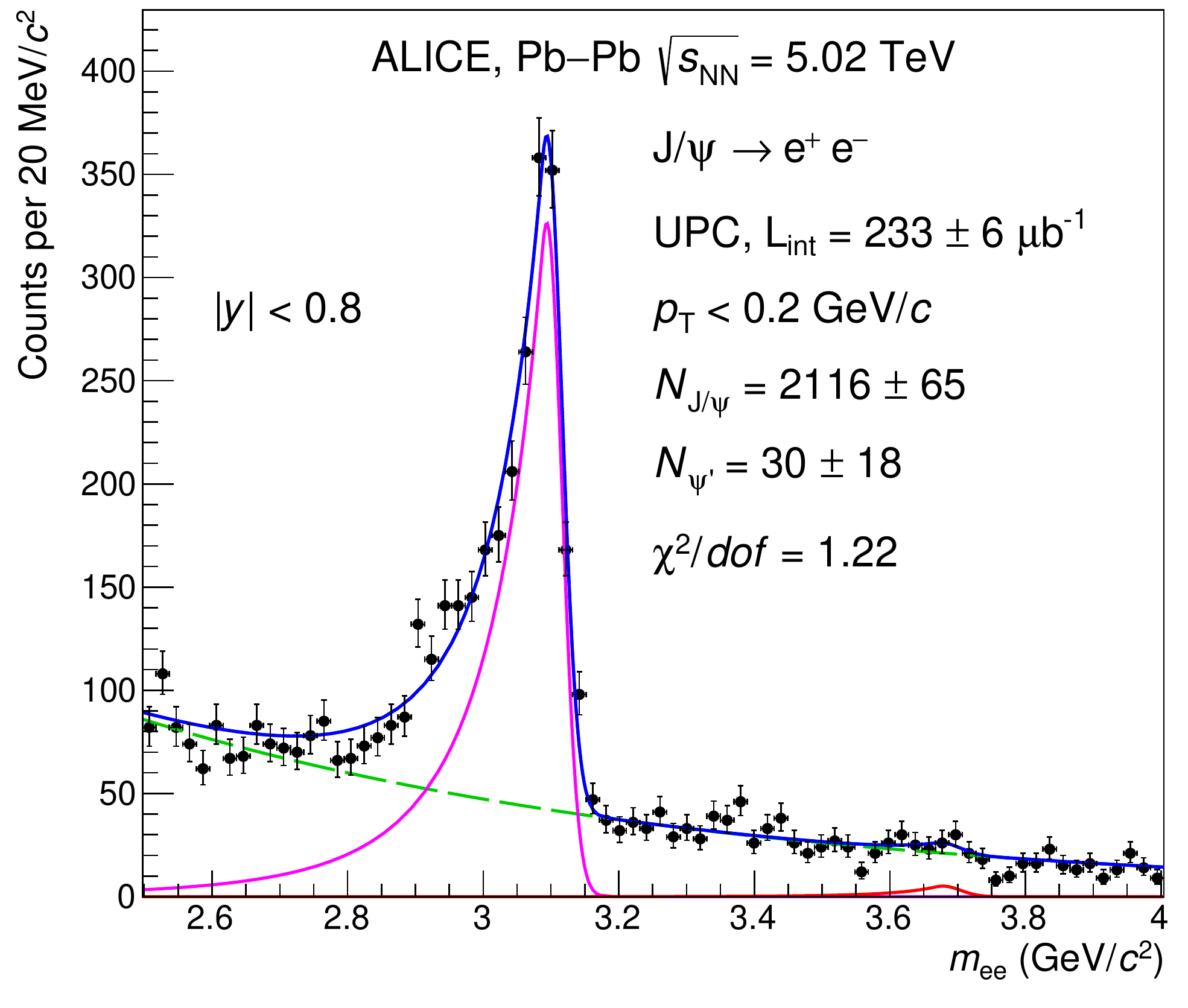}
\includegraphics[width=0.49\textwidth]{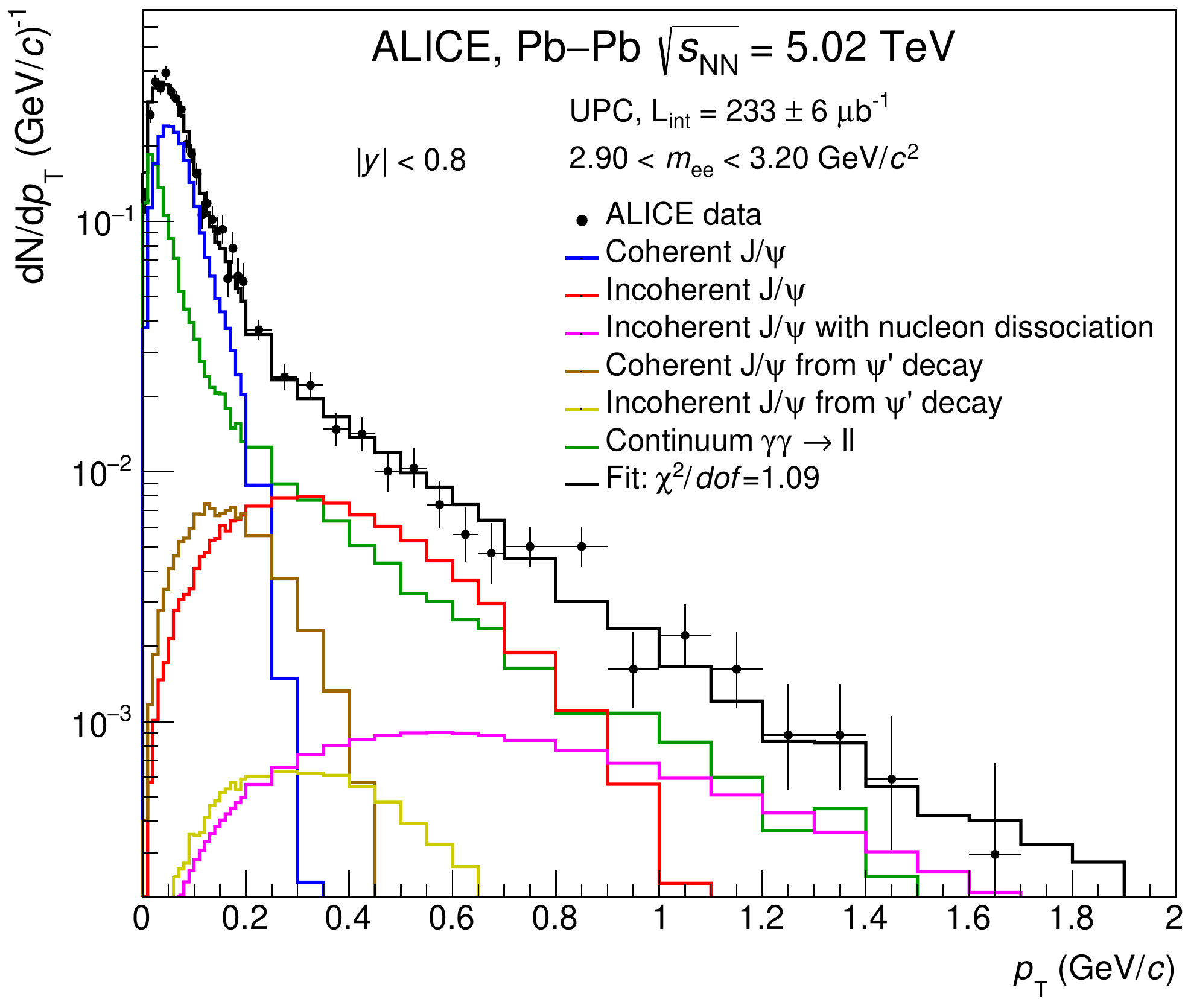}}

\caption{Left: Invariant mass distribution of \lplp\ pairs. The dashed green line corresponds to the background. The solid magenta and red lines correspond to Crystal Ball functions representing the \Jpsi\ and \Ppsi\ signal, respectively. The solid blue line corresponds to the sum of background and signal functions. Right: Transverse momentum distribution of \Jpsi\ candidates in the range quoted in the figure (around the \Jpsi\ nominal mass).}
\label{fig:MassPtJpsiMuEl}
\end{figure}

The invariant mass distributions for dimuon and dielectron pairs reconstructed in the full rapidity range are shown in Fig.~\ref{fig:MassPtJpsiMuEl}, left.
The inclusive \Jpsi~yields are obtained by fitting the invariant mass distributions with an exponential function describing the underlying continuum and two Crystal Ball functions to describe the \Jpsi~and \Ppsi~signals. The \Jpsi\ pole mass and width were left free, while the tail parameters  ($\alpha$ and $n$) in the Crystal Ball function were fixed to the values obtained in MC simulations in order to gain higher stability of the fits. In the case of the \Ppsi\ signal, all the Crystal Ball parameters were fixed to the values obtained in MC simulations. 

The raw inclusive \Jpsi~yields obtained from invariant mass fits contain contributions from the coherent and incoherent \Jpsi~photoproduction that can be separated via the analysis of the transverse momentum spectra. The inclusive \pt~distributions for \mumu\  and \elel\ candidates around the \Jpsi~mass are shown in the right panels of Fig.~\ref{fig:MassPtJpsiMuEl}. These distributions are fitted with MC templates produced using STARlight, followed by full detector simulation and reconstruction, corresponding to different production mechanisms: coherent and incoherent \Jpsi, feed-down \Jpsi\ from decays of coherent and incoherent \Ppsi\ and the dilepton continuum from the $\gamma \gamma \rightarrow  l^{+}l^{-}$ process. 
Incoherent \Jpsi\ production with nucleon dissociation (or dissociative \Jpsi) is also taken into account to describe the high-\pt~tail with the template based on the H1 parametrization~\cite{Alexa:2013xxa}. Normalization of feed-down \Jpsi~from coherent and incoherent \Ppsi~decays is constrained to the normalization of primary \Jpsi~templates according to the feed-down fractions extracted as described below. The normalization of the dilepton continuum from the $\gamma \gamma \rightarrow  l^{+}l^{-}$ process is fixed by the results for the background description of the invariant mass fits. The combinatorial background, estimated by considering the distribution of like-sign candidates, is found to be negligible in the \Jpsi~mass region.

The templates are fitted to the data leaving the normalization free for coherent \Jpsi, incoherent \Jpsi\ and dissociative \Jpsi\ production. 
The extracted incoherent \Jpsi\ fraction $f_{\rm I} = \frac{N^{\rm incoh}}{N^{\rm coh}}$ for $\pt<0.2$ GeV/$c$\ is $4.7 \pm 0.3\%$ $(5.0 \pm 0.5)\%$ for the \mumu(\elel) decay channel. The quoted fractions include the contribution of incoherent \Jpsi\ with nucleon dissociation.

\begin{figure}
\centering
\includegraphics[width=0.49\textwidth]{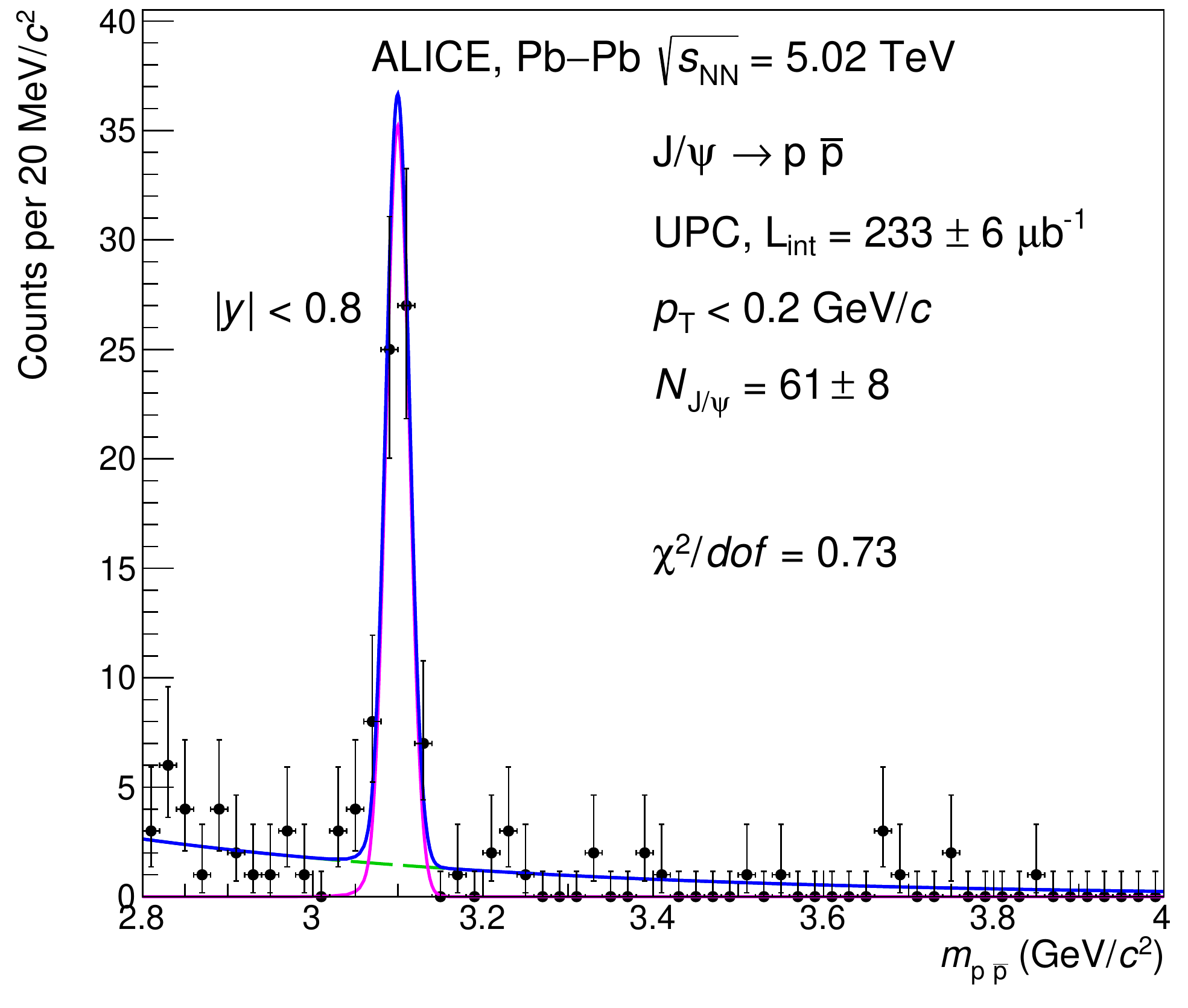}
\includegraphics[width=0.49\textwidth]{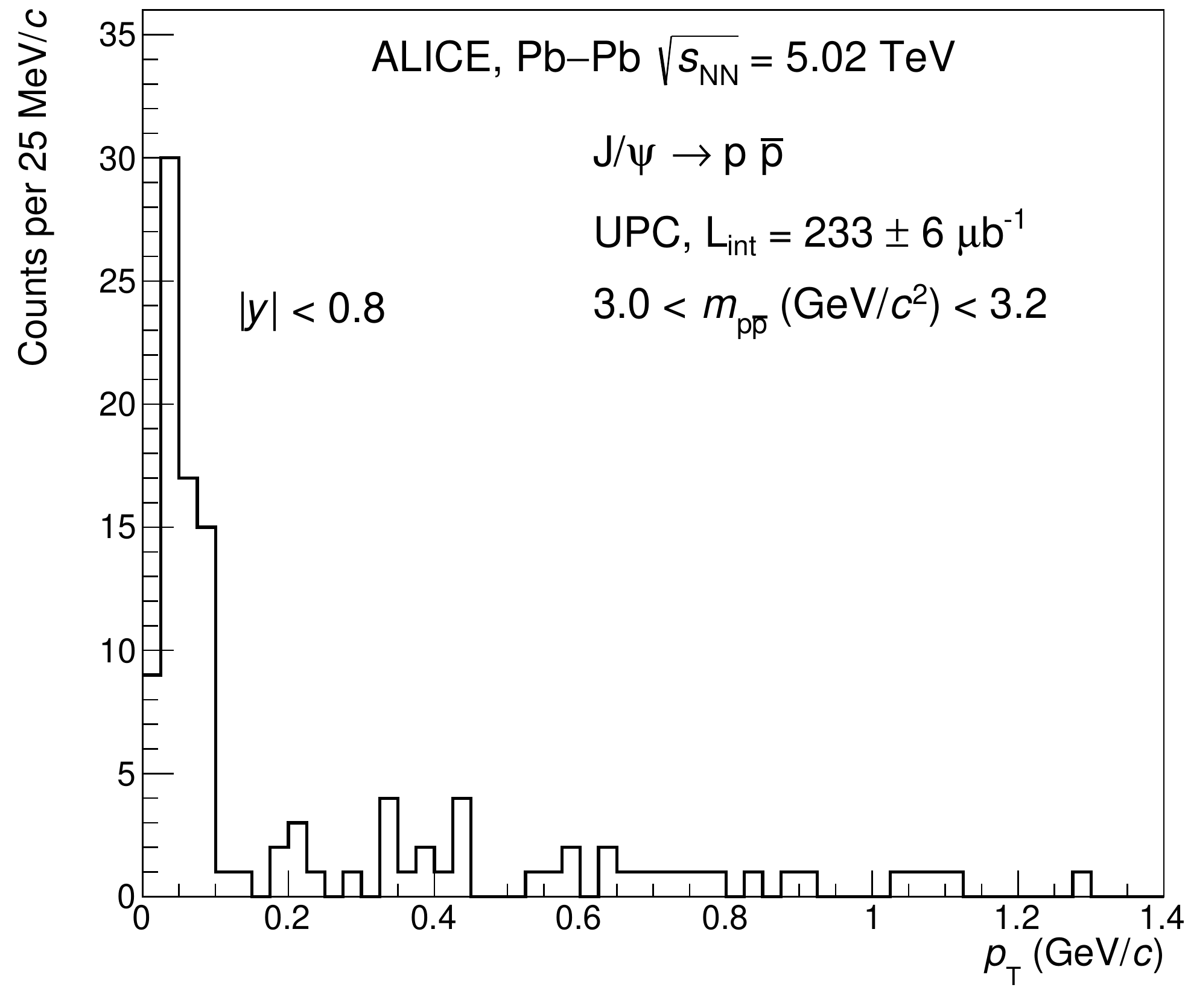}
\caption{Left: Invariant mass distribution of \pbp\ pairs. The dashed green line corresponds to the background description. The solid magenta and red lines correspond to Crystal Ball functions representing the \Jpsi\ and \Ppsi\ signals, respectively. The solid blue line corresponds to the sum of background and signal functions. Right: Transverse momentum distribution of \Jpsi\ candidates in the range quoted in the figure (around the \Jpsi\ nominal mass).}
\label{fig:MassPtJpsiPP}
\end{figure}

The invariant mass and the \pt~distributions for the $\rm{J}/\psi \rightarrow {\rm p \bar p}$ decay channel are shown in Fig.~\ref{fig:MassPtJpsiPP}.
The data sample obtained in this channel is too small to fit the \pt~distribution with MC templates. However, since the difference in resolution of \pt~shapes of the coherent or incoherent MC samples for the \pbp\ and \mumu\ channels is negligible, one can expect the $f_{\rm I}$ fraction to be the same. This is due to the fact that neither the \mumu\ nor the \pbp\ channels suffer from bremsstrahlung. This is not the case for dielectrons where bremsstrahlung induces the large difference in the mass and momentum resolution which affect the templates and consequently the $f_{\rm I}$ fraction.

\begin{figure}
\centering
\includegraphics[width=0.49\textwidth]{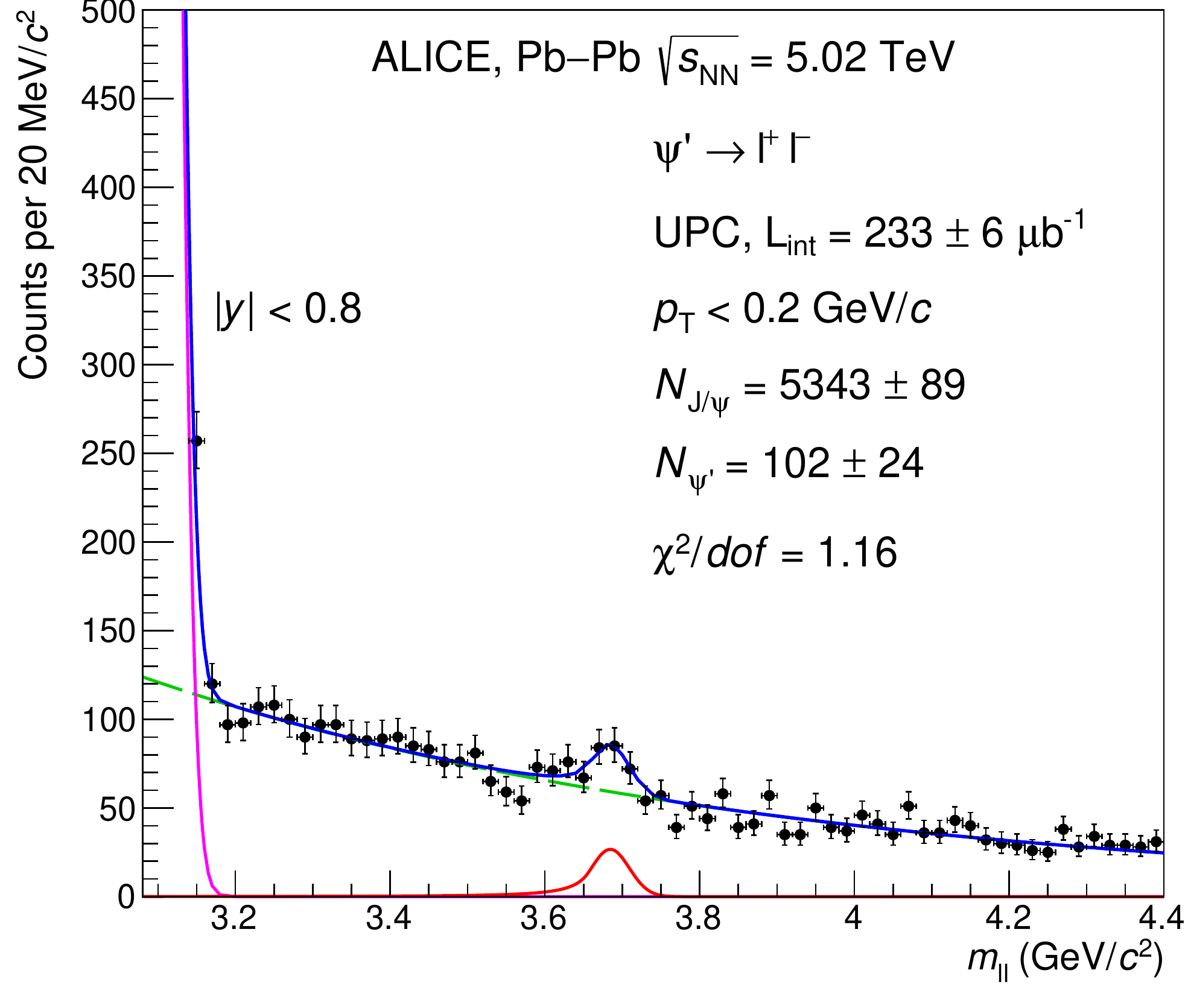}
\includegraphics[width=0.49\textwidth]{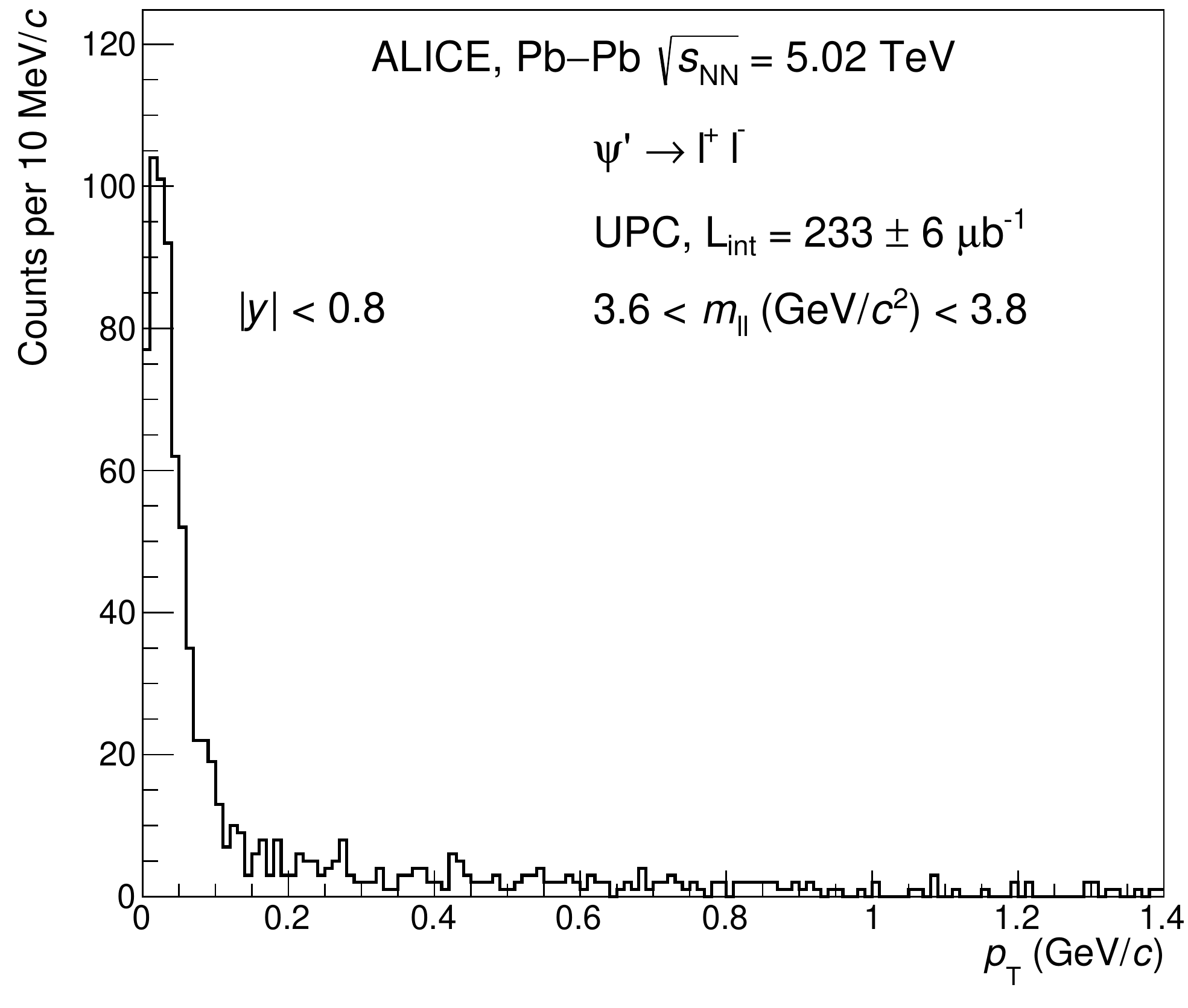}

\caption{Left: Invariant mass distribution of \lplp\ pairs. The dashed green line corresponds to the background description. The solid magenta and red line correspond to Crystal Ball functions representing the \Jpsi\ and \Ppsi\ signals, respectively. The solid blue line corresponds to the sum of background and signal functions. Right: Transverse momentum distribution of \Ppsi\ candidates in the mass range quoted in the figure (around the \Ppsi\ mass).}
\label{fig:MassPtPsi2s2T}
\end{figure}

As one can see in Fig.~\ref{fig:MassPtJpsiMuEl}, the \Ppsi~yields in the \mumu\ and \elel\ channels are small and lying on top of a significant background. In order to increase the significance of the \Ppsi\ signal and to reduce the statistical uncertainty, the \Ppsi\ yield is extracted from the merged \lplp\ sample with significance $\approx3$.
Fig.~\ref{fig:MassPtPsi2s2T} shows the merged dilepton mass spectrum together with the \pt\ distribution of the dilepton candidates in the invariant mass range under the \Ppsi\ mass peak.
The fit to the invariant mass distribution is performed in the same way as described before.

\begin{figure}
\centering
\subfigure[\Ppsi $\rightarrow$ \mmpp]
{\label{fig:MassPtPsi2s4T_a}\includegraphics[width=0.49\textwidth]{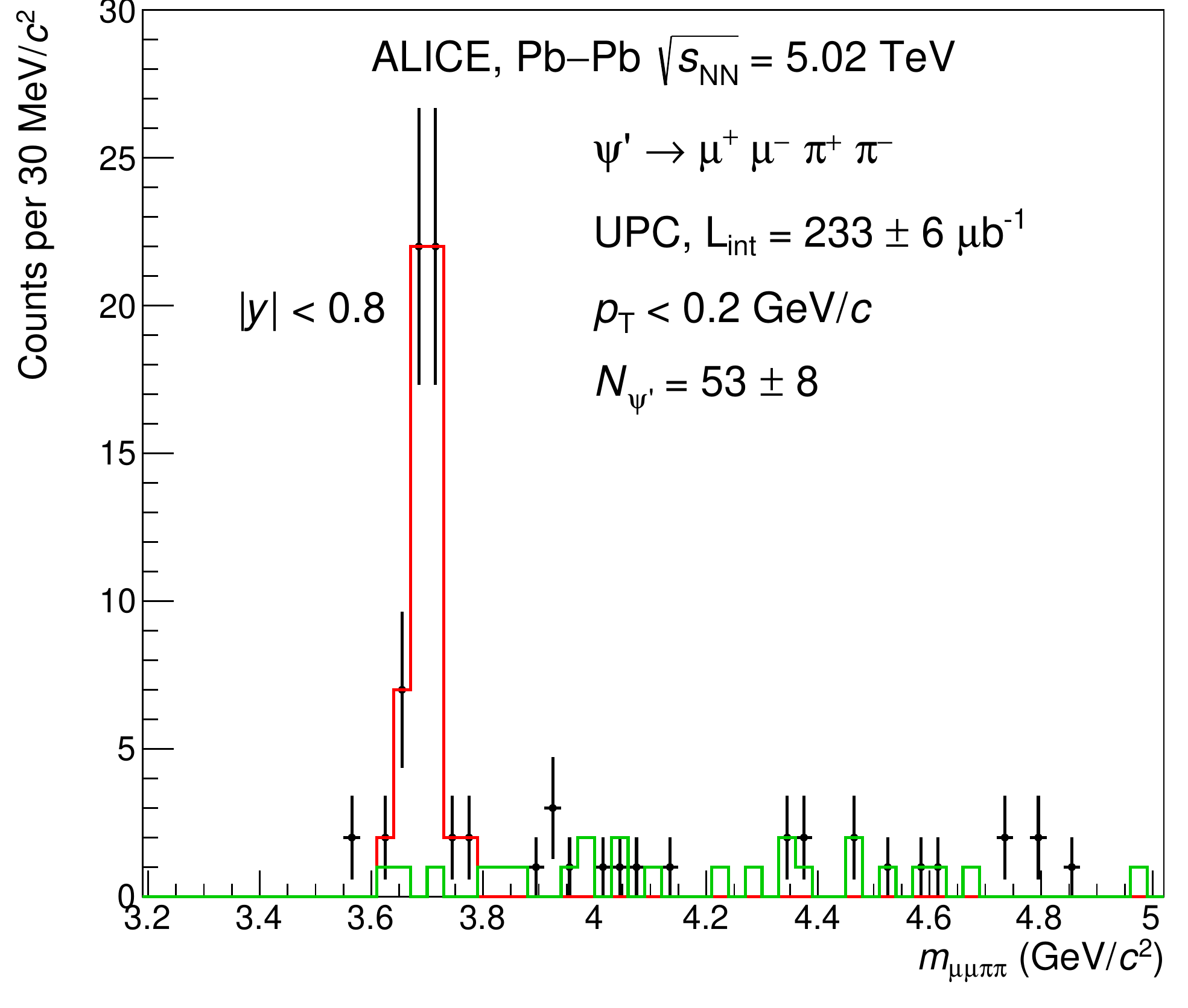}
\includegraphics[width=0.49\textwidth]{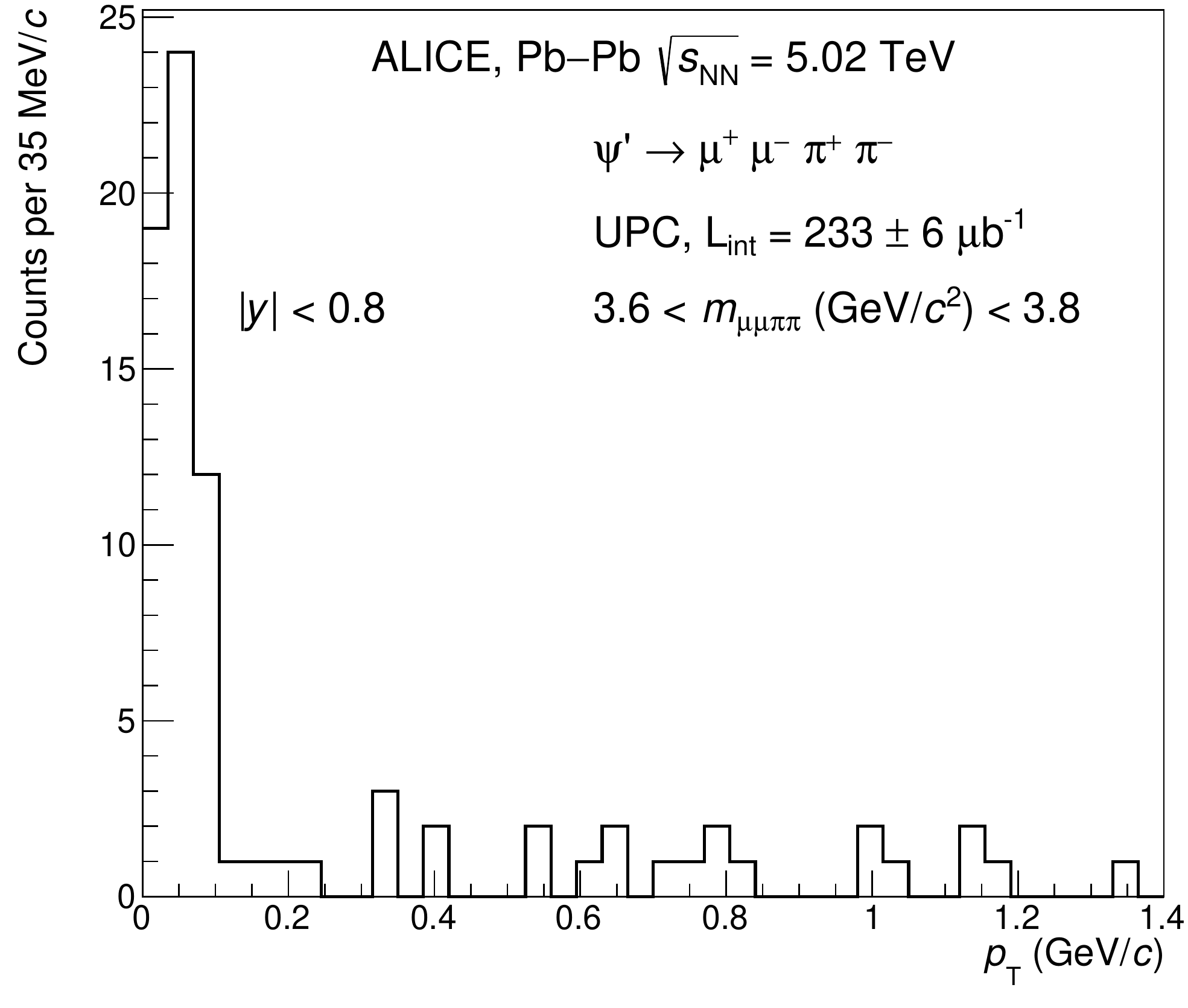}}
\subfigure[\Ppsi $\rightarrow$ \eepp]
{\label{fig:MassPtPsi2s4T_b}\includegraphics[width=0.49\textwidth]{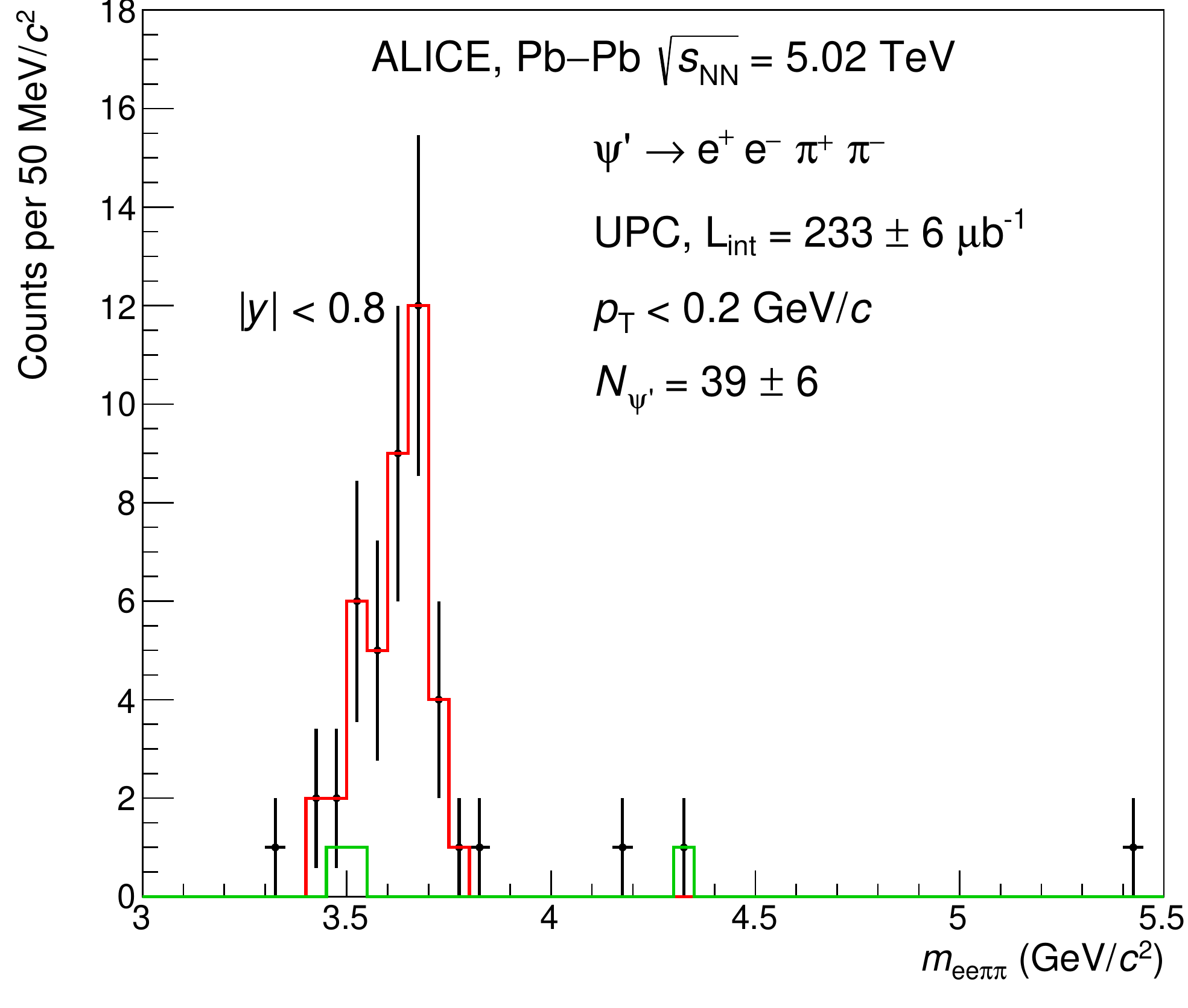}
\includegraphics[width=0.49\textwidth]{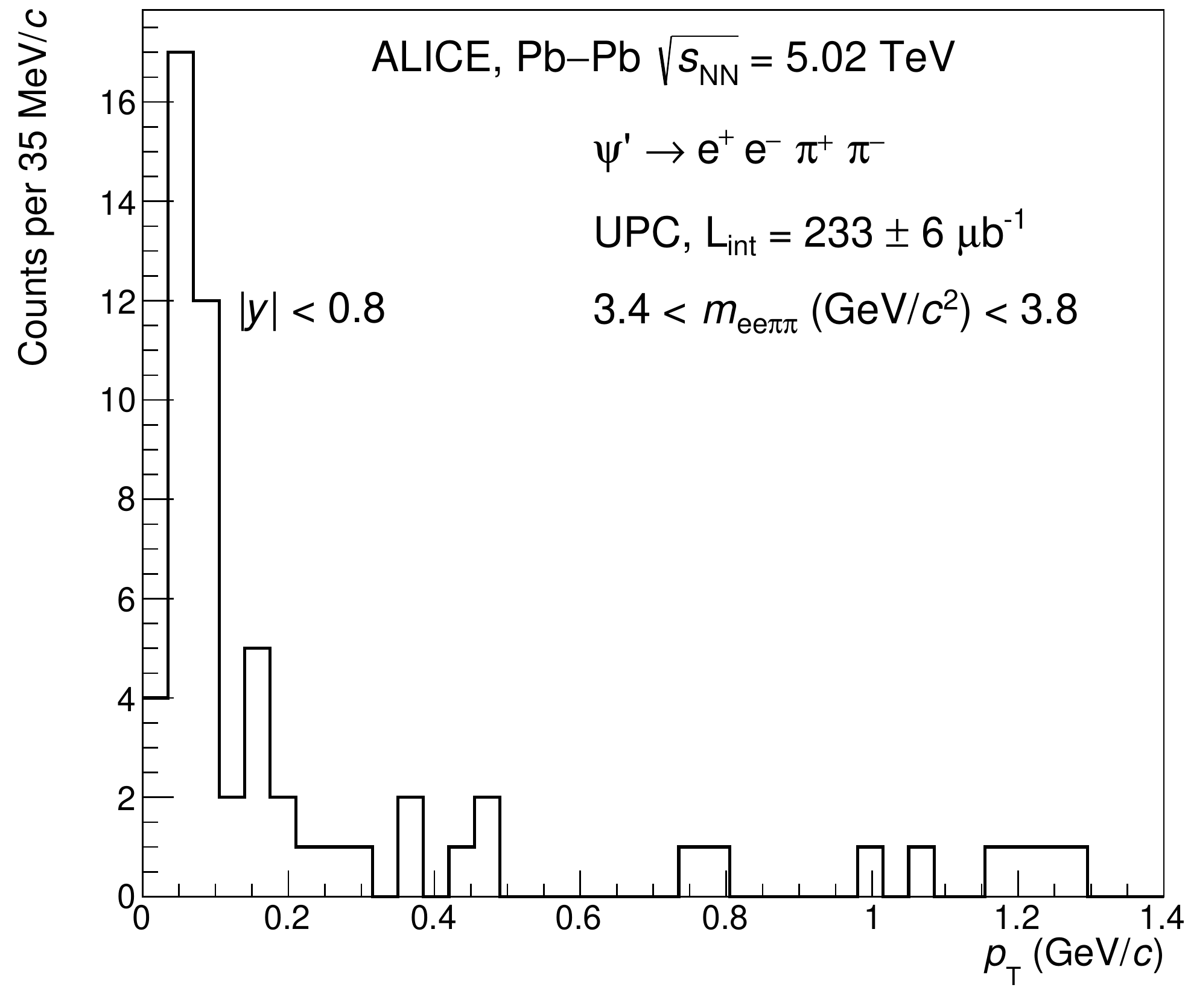}}

\caption{Left: Invariant mass distribution for \Ppsi\ $\rightarrow$ \mmpp\ (upper panel) and  \Ppsi\ $\rightarrow$ \eepp\ (bottom panel). The green line shows the wrong sign four-track events. The red line shows the \Ppsi\ signal as described in the text. Right: Transverse momentum distribution of \Ppsi~candidates in the mass range quoted in the figure (around the \Ppsi\ nominal mass).}
\label{fig:MassPtPsi2s4T}
\end{figure}

Fig.~\ref{fig:MassPtPsi2s4T} shows the invariant mass (left) and the \pt~distribution (right) for \Ppsi\ $\rightarrow$ \mmpp\ and \Ppsi\ $\rightarrow$ \eepp\ quadruplets. Distributions include combinatorial background. A coherent peak is clearly visible at low \pt.
The signal extraction in the \mmpp\ and \eepp\ channel is straight-forward since the signal is very clean. 
The number of candidates is extracted by summing the bin contents in the mass interval $\rm 3.6<m_{\mu\mu\pi\pi}<3.8$\,GeV/$c^2$ and $\rm 3.4<m_{ee\pi\pi}<3.8$\,GeV/$c^2$. The number of candidates with wrong-sign combinations in the same mass interval, representing the level of background, is subtracted afterwards.

The incoherent contamination of the \Ppsi\ sample is estimated as follows. The incoherent-to-coherent photoproduction cross section ratio is expected to be similar for 1S and 2S charmonium states~\cite{Ducati:2013bya,Klein:2016yzr}. Due to lack of model calculations for the incoherent \Ppsi\ cross section in UPCs at \fivenn, predicted incoherent-to-coherent cross section ratios for \Jpsi\ from Refs.~\cite{Klein:2016yzr, Cepila:2017nef, Guzey:2018tlk} are used as an estimate of the incoherent-to-coherent cross section ratio for \Ppsi. The factor $f_{\rm I} = \frac{N^{\rm incoh}}{N^{\rm coh}} \approx 6\%$ is extracted from the predicted
cross section ratios, corrected for acceptance and efficiency of coherent and incoherent \Ppsi\ states. The difference in the predicted incoherent-to-coherent cross section ratios is used as an estimate of the systematic uncertainty.

The raw \Jpsi\ yields contain a significant feed-down contribution originating from decays $\psi'\ \rightarrow {\rm J}/\psi + {\rm anything}$, dominated by the \Ppsi\ $\rightarrow$ \Jpsi\ + $\pi^{+} \pi^{-}$ and \Ppsi\ $\rightarrow$ \Jpsi + $\pi^{0} \pi^{0}$ decay channels. The feed-down fraction $f_{\rm D} = \frac{N({\rm FeedDown})}{N({\rm primary})}$ can be extracted from the ratio of raw \Jpsi\ and \Ppsi\ yields:  
\begin{equation}
R_N = \frac{N_{\psi'}}{N_{J/\psi}} = 0.0170 \pm 0.0024 (0.0184 \pm 0.0030), 
\label{eq:FeedDownRN}
\end{equation}
for \mumu(\elel).
The raw \Ppsi\ and \Jpsi\ yields in this ratio contain contributions both from coherent and incoherent photoproduction. However, according to the \pt\ fits, the fraction $f_{\rm I}$ does not exceed 6\% and, according to STARlight, the fraction of the incoherent contributions is expected to be similar in the \Ppsi\ and \Jpsi\ yields. The $R_N$ ratio can therefore be considered as a good estimate of the ratio of coherent \Jpsi\ and \Ppsi\ yields, since the incoherent fractions largely cancel in the ratio. The $f_{\rm D}$ ratio can be expressed via the measured $R_N$ ratio:
\begin{equation}
\left(
\frac{1}{f_{\rm D}}+1 
\right)^{-1}
= 
\frac{N_{{\rm J}/\psi}^{\rm feed-down}}{N_{{\rm J}/\psi}}
=
\frac{ ({\rm BR} \times   \epsilon)_{\psi'\rightarrow {\rm J}/\psi\pi^+\pi^-\rightarrow l^+l^-\pi^+\pi^-}+
({\rm BR} \times   \epsilon)_{\psi'\rightarrow {\rm J}/\psi\pi^0\pi^0\rightarrow l^+l^-\pi^0\pi^0}
}{({\rm BR}\times \epsilon)_{\psi' \to l^+l^-}}
\times R_{N}
,
\label{eq:FeedDownDT}
\end{equation}
where $({\rm BR} \times \epsilon)$ in the corresponding channels denote products of world-average branching ratios~\cite{Zyla:2020zbs} and the product of acceptance and efficiency of measuring exactly two leptons. The $f_{\rm D}$~fractions of
$3.5\% \pm 0.5\%$ and $4.3\% \pm 0.7\%$ are obtained in the \mumu  and  in \elel\ channel respectively, with the uncertainty being the quadratic sum of statistical and systematic uncertainties, where the statistical uncertainty dominates.  
The systematic uncertainty includes contributions from the \Jpsi\ and \Ppsi\ signal extraction and the branching ratios.

\FloatBarrier
\section{Results}

The coherent vector meson differential cross section is given by:
\begin{equation}
\frac{\rm {d}\sigma^{coh}_{\rm VM}}{\rm{d} y} = \frac{N^{\rm coh}_{\rm VM}}{\epsilon_{\rm VM} ~\times~ \epsilon_{\rm veto}^{\rm pileup} ~\times~ \epsilon_{\rm veto}^{\rm EMD} ~\times~ \rm{BR} ~\times~ \mathcal{L}_{\rm int} ~\times~ \Delta y}
\label{eq:Xsection}
\end{equation}
\noindent where
\begin{equation}
N^{\rm coh}_{J/\psi} = \frac{N_{\rm yield}}{1+f_{\rm I}+f_{\rm D}}
\label{eq:YieldJpsi}
\end{equation}
\noindent and
\begin{equation}
N^{\rm coh}_{\psi'} = \frac{N_{\rm yield}}{1+f_{\rm I}}.
\label{eq:YieldPsi2s}
\end{equation}

The raw \Jpsi\ and \Ppsi\ yield values, reconstruction efficiencies $\epsilon$, fractions corresponding to incoherent $f_{\rm I}$ and feed-down $f_{\rm D}$ contamination as well as coherent cross sections with statistical and systematic uncertainties are summarized in Table~\ref{tab:CrossSectionJpsi} and Table~\ref{tab:CrossSectionPsi2s}. Further components in the formula include integrated luminosity $\mathcal{L}_{\rm int}$, branching ratio of the decay BR, veto efficiencies corresponding to the pile-up $\epsilon_{\rm veto}^{\rm pileup}$ and electromagnetic dissociation $\epsilon_{\rm veto}^{\rm EMD}$ and the rapidity interval $\Delta \rm y$. The associated systematic uncertainties are briefly described in the following.

\begin{table*}[htbp]
\caption{Raw \Jpsi\ yields, $\epsilon$, $f_{\rm D}$ and $f_{\rm I}$ fractions and coherent \Jpsi\ cross sections per decay channel}
\centering
\begin{tabular}{lcccccc}
\toprule
Decay & $|y|$ & $N_{J/\psi}$ & $\epsilon$ & $f_{\rm D}$ & $f_{\rm I}$ & $\rm {d}\sigma^{coh}_{J/\psi}/\ \rm{d} y$ (mb) \\ 
\midrule
\mumu & (0.00, 0.80) & $3120 \pm 61$ & 0.037 & 0.035 & $0.047 \pm 0.003$ & $4.12 \pm 0.08 \rm (stat.) \pm 0.23(syst.) $ \\
\addlinespace
\mumu & (0.00, 0.15) & $1027 \pm 35$ & 0.064 & 0.035 & $0.047 \pm 0.003$ & $4.23 \pm 0.15 \rm (stat) \pm 0.24(syst) $ \\ 
\mumu & (0.15, 0.35) & $1083 \pm 36$ & 0.051 & 0.035 & $0.047 \pm 0.003$ & $4.22 \pm 0.14 \rm (stat) \pm 0.23(syst) $ \\ 
\mumu & (0.35, 0.80) & $976 \pm 33$ & 0.022 & 0.035 & $0.047 \pm 0.003$ & $3.85 \pm 0.13 \rm (stat) \pm 0.21(syst) $ \\ 
\addlinespace
\elel & (0.00, 0.80) & $2116 \pm 65$ & 0.025 & 0.043 & $0.050 \pm 0.005$ & $4.05 \pm 0.13 \rm (stat.) \pm 0.24(syst.) $ \\
\addlinespace
\elel & (0.00, 0.15) & $683 \pm 33$ & 0.046 & 0.043 & $0.050 \pm 0.005$ & $3.83 \pm 0.19 \rm (stat) \pm 0.23(syst) $ \\ 
\elel & (0.15, 0.35) & $743 \pm 34$ & 0.034 & 0.043 & $0.050 \pm 0.005$ & $4.20 \pm 0.19 \rm (stat) \pm 0.25(syst) $ \\ 
\elel & (0.35, 0.80) & $643 \pm 31$ & 0.014 & 0.043 & $0.050 \pm 0.005$ & $3.90 \pm 0.19 \rm (stat) \pm 0.23(syst) $ \\ 
\addlinespace
\pbp & (0.00, 0.80) & $61 \pm 8$ & 0.023 & 0.035 & $0.047 \pm 0.003$ & $3.73 \pm 0.51 \rm (stat.) \pm 0.28(syst.) $ \\ 
\bottomrule
\end{tabular}
\label{tab:CrossSectionJpsi}
\end{table*}

\begin{table*}[htbp]
\caption{Raw \Ppsi\ yields, $\epsilon$, $f_{\rm I}$ fractions and coherent \Ppsi\ cross sections per decay channel}
\centering
\begin{tabular}{lccccc}
\toprule
Decay & $|y|$ & $N_{\psi'}$ & $\epsilon$ & $f_{\rm I}$ & $\rm {d}\sigma^{coh}_{\psi'}/ \rm{d} y$ (mb) \\ 
\midrule
\mmpp & (0.0, 0.8) & $53 \pm 8$ & 0.0103 & 0.057 & $0.77 \pm 0.11 \rm (stat.) \pm 0.09(syst.) $ \\
\addlinespace
\eepp & (0.0, 0.8) & $39 \pm 6$ & 0.0068 & 0.064 & $0.86 \pm 0.15 \rm (stat.) \pm 0.11(syst.) $ \\
\addlinespace
\lplp & (0.0, 0.8) & $102 \pm 24$ & 0.0324 & 0.063 & $0.61 \pm 0.15 \rm (stat.) \pm 0.10(syst.) $ \\
\bottomrule
\end{tabular}
\label{tab:CrossSectionPsi2s}
\end{table*}

\begin{figure}
\centering
\includegraphics[width=0.49\textwidth]{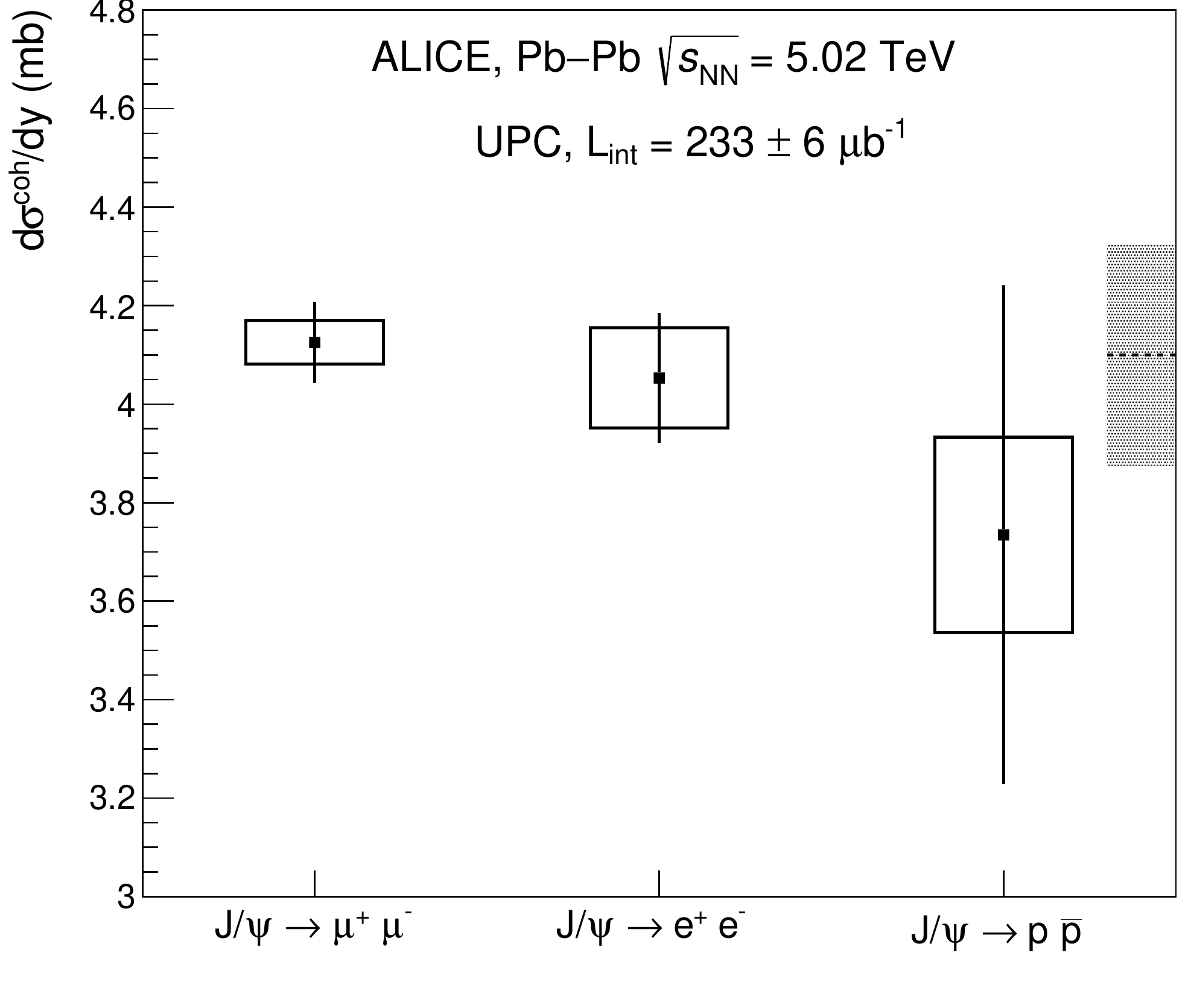}
\includegraphics[width=0.49\textwidth]{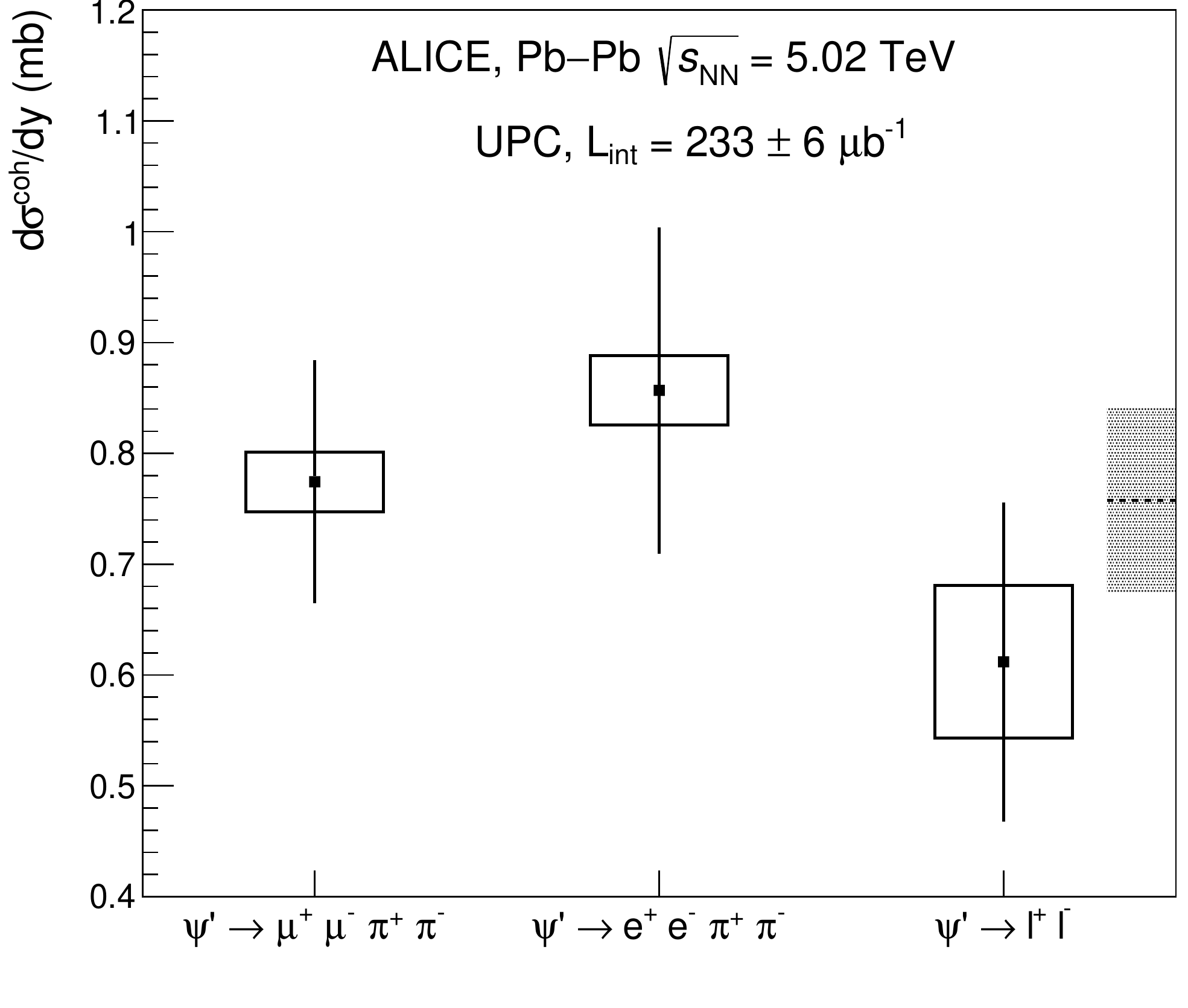}
\caption{Measured differential cross section of coherent \Jpsi\ (left) and \Ppsi\ (right) photoproduction in \PbPb\ UPCs in $|y|<0.8$. The points show the measurements for different decay channels. The error bars (boxes) represent the statistical (decay channel uncorrelated systematic) uncertainty. The gray box shows the average value (dashed line) and correlated systematic uncertainty.}
\label{fig:XsectionChannels}
\end{figure}

The offline AD and V0 vetoes decreases the \Jpsi\ (\Ppsi) yield by 8\% (16\%) and also results in a lower $\epsilon_{{\rm veto}}^{{\rm pileup}}$ efficiency. These two effects do not cancel-out completely.
A residual discrepancy of 3\% (10\%) in the cross section of \Jpsi\ (\Ppsi) is included in the systematic uncertainty.  

The uncertainty related to the evaluation of the incoherent contamination comes from the shape of the continuum  template. By default the $\gamma\gamma \to l^+ l^-$ template from the STARlight MC simulation is used. Alternative way would be to use a data-driven template generated from the side bands of the invariant mass distribution of the \Jpsi. These events are, in contrast to the pure MC, supposed to include the same background as under the \Jpsi\ peak. By comparing the $f_{\rm I}$ fraction obtained with the side bands method and the STARlight template, a 0.8 (0.5\%) uncertainty of the cross section for the \mumu\ (\elel) channel is found.

A systematic uncertainty of the tracking efficiency of 2\% per track is estimated by comparing, in data and in MC, the ITS (TPC) hit matching efficiency to the tracks reconstructed with TPC (ITS) hits only. This leads to a 2.8\% (4\%) systematic uncertainty for two-track (four-track) channels.

For the signal extraction in the \Jpsi\ analysis the goodness of the description of the \Jpsi\ signal by the Crystal Ball function is checked. The yield from the fit is compared to the number of events computed by bin counting in the peak region with the background contribution subtracted using the exponential background shape from the fit. Half differences of 0.3\%, 2.4\%, 0.6\% were assigned as the systematic uncertainty in the muon, electron and proton channel, respectively.

Another contribution to the signal extraction uncertainties is the difference between the function used in the fit and the true shape of the background. It is estimated by varying the fit range. A systematic uncertainty of 0.4\% (0.3\%) is determined for the \mumu\ (\elel) decay channel of the \Jpsi meson. The uncertainty in the background subtraction rises rapidly as the signal-to-background ratio drops. A similar study for the \lplp\ decay of the \Ppsi\ meson results in a 10\% systematic uncertainty.

The uncertainty associated to the determination of the trigger efficiency of the SPD chips is obtained by changing the requirements on the probe tracks used in the data-driven method. Variations include the running conditions, the maximum amount of activity allowed in the event, and the definition of tracks accepted in the efficiency computation. This uncertainty amounts to 1\%.

The uncertainty of the TOF trigger efficiency due to the spread of the arrival times of various particle species to TOF is evaluated as 0.5\% per track (1\% in total). The uncertainty in case of four track decays of \Ppsi~applies only for the lepton tracks since the low-momentum pions do not reach the TOF detector.

In the ${\rm J}/\psi \rightarrow \rm p \overline{p}$ analysis, at least one track is required to have proton PID from the TOF. Comparing the efficiency of the track matching to TOF in data and MC samples, a $10\%$ disagreement is found. The matching efficiency from MC is used and a half-difference of 5\% as an additional systematic uncertainty for the $\rm p \overline{p}$ channel is assigned.

Tables~\ref{tab:SystematicsJpsi} and~\ref{tab:SystematicsPsi2s} show the uncertainties for each source and channel separately as well as quadratic sums of the channel-correlated and uncorrelated sources.

The signal extraction, incoherent contamination and branching ratio are considered as channel-uncorrelated sources of systematic uncertainties. The other sources are fully correlated and are the same for all channels. In the case of the \Ppsi,~the four track channels have an extra ITS-TPC matching uncertainty for the pion tracks which is not correlated with the \lplp\ channel, thus it is quoted separately.

\begin{table*}[htbp]
\caption{Sources of systematic uncertainty for the coherent \Jpsi\ cross section measurements per decay channel in percent}
\centering
\begin{tabular}{l ccc}
\toprule
 & $\rm J/\psi \rightarrow \mu^{+} \mu^{-}$ & ${\rm J}/\psi \rightarrow e^{+} e^{-}$ & $\rm J/\psi \rightarrow p \overline{p}$  \\
\midrule
Signal Extraction &  0.5 &  2.4 &  0.7 \\
Incoherent contamination & 0.8 & 0.5 & 0.8  \\
Branching ratio & 0.5 & 0.5 & 1.4\\
TOF matching & -- & -- & 5.0\\
ITS-TPC matching &  2.8 &  2.8 &  2.8 \\
AD and V0 veto &  3.0 &  3.0 &  3.0 \\
SPD trigger efficiency &  1.0 &  1.0 &  1.0 \\
TOF trigger efficiency &  0.7 &  0.7 &  0.7 \\
Luminosity & 2.7 & 2.7 & 2.7\\
EMD correction & 2.0 & 2.0 & 2.0 \\
Feed down & 0.6 & 0.6  & 0.6 \\
\addlinespace
\midrule
Channel uncorrelated &  1.1 &  2.5 &  5.3 \\
\addlinespace
Channel correlated &  5.5 &  5.5 &  5.5\\
\bottomrule
\end{tabular}
\label{tab:SystematicsJpsi}
\end{table*}

\begin{table*}[htbp]
\caption{Sources of systematic uncertainty for the coherent \Ppsi\ cross section measurements per decay channel in percent}
\centering
\begin{tabular}{l ccc}
\toprule
 & $\psi' \rightarrow \mu^{+} \mu^{-} \pi^{+} \pi^{-}$ & $ \psi' \rightarrow e^{+} e^{-} \pi^{+} \pi^{-}$  &  $  \psi' \rightarrow l^{+} l^{-}$  \\
\midrule
Signal Extraction &  1.0 &  2.0 &  10.0 \\
Incoherent contamination & 1.4 & 1.8 & 1.8  \\
Branching ratio & 1.5 & 1.5 & 4.8\\
ITS-TPC matching pions &  2.8 &  2.8 & -- \\
ITS-TPC matching leptons &  2.8 &  2.8 &  2.8 \\
AD and V0 veto &  10.0 &  10.0 &  10.0 \\
SPD trigger efficiency &  1.0 &  1.0 &  1.0 \\
TOF trigger efficiency &  0.7 &  0.7 &  0.7 \\
Luminosity & 2.7 & 2.7 & 2.7\\
EMD correction & 2.0 & 2.0 & 2.0 \\
\addlinespace
\midrule
Channel uncorrelated &  3.5 &  5.8 &  11.2 \\
\addlinespace
Channel correlated &  11.0 &  11.0 &  11.0\\
\bottomrule
\end{tabular}
\label{tab:SystematicsPsi2s}
\end{table*}

The \Jpsi\ and \Ppsi\ cross sections for various decay channels computed using Eq.~\ref{eq:Xsection} are shown in Fig.~\ref{fig:XsectionChannels}. The mean values of the \Jpsi\ and \Ppsi\ cross sections are obtained as a weighted average of the cross sections per decay channel with weights corresponding to the inverse of the quadratic sum of statistical and channel-uncorrelated systematic uncertainties. 
The cross section value averaged over the three decay channels of the coherent \Jpsi~photoproduction measurements is: 
\begin{equation}
\frac{{\rm d}\sigma^{\rm coh}_{J/\psi}}{{\rm d} y} = 4.10~\pm 0.07\rm(stat.) \pm 0.23 \rm(syst.)  mb.
\end{equation}
The cross section value averaged over the three channels of coherent \Ppsi~photoproduction measurements is: 
\begin{equation}
\frac{{\rm d}\sigma^{\rm coh}_{\psi'}}{{\rm d} y} = 0.76~\pm 0.08\rm(stat.) \pm 0.09 \rm(syst.)  mb.
\end{equation}

The ratio of the 2S to 1S charmonium states is:
\begin{equation}
\frac{\frac{\sigma^{\rm coh}_{\psi'}}{{\rm d} y}}{\frac{\sigma^{\rm coh}_{J/\psi}}{{\rm d} y}} = 0.18~\pm 0.0185\rm(stat.) \pm 0.028 \rm(syst.) \pm 0.005 \rm(BR).
\end{equation}

Many systematic uncertainties of the \Jpsi\ and \Ppsi\ cross section measurements are correlated and cancel in the cross section ratio. Since the analysis relies on the same data sample and on the same trigger, the systematic uncertainties of the luminosity evaluation, trigger efficiency, EMD correction and ITS-TPC matching of leptons were considered as fully correlated. The AD and V0 offline veto uncertainty is partially correlated, so the difference of the uncertainties for \Ppsi\ and \Jpsi\ is taken into account in the uncertainty of the ratio. The systematic uncertainties connected to the signal extraction, incoherent contamination and the branching ratio are considered uncorrelated between the two measurements. 
The dominant uncertainty comes from the uncorrelated part of the AD and V0 veto uncertainty for \Ppsi.

\FloatBarrier
\section{Discussion}

Figure~\ref{fig:XsectionPredictions} shows the rapidity-differential cross section of the coherent photoproduction of \Jpsi\ and \Ppsi~vector mesons in Pb–Pb UPCs including previous ALICE measurements of \Jpsi\ at forward rapidity~\cite{Acharya:2019vlb}. At midrapidity, \Jpsi\ measurements performed in absolute rapidity ranges are shown at positive rapidities and reflected into negative rapidities. The ALICE measurements are compared to several models which are discussed in the following:

\begin{figure}
\centering
\includegraphics[width=0.49\textwidth]{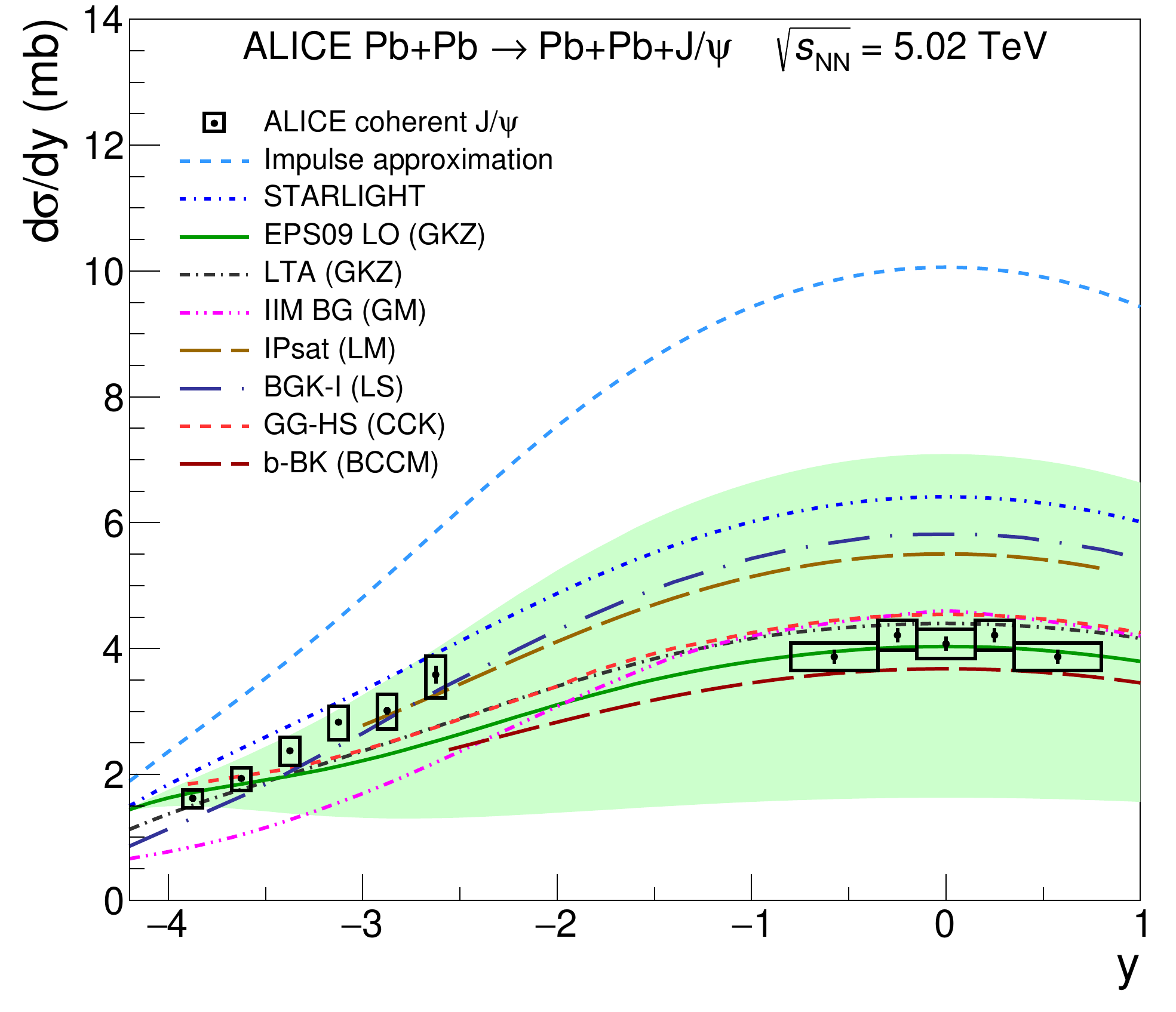}\includegraphics[width=0.49\textwidth]{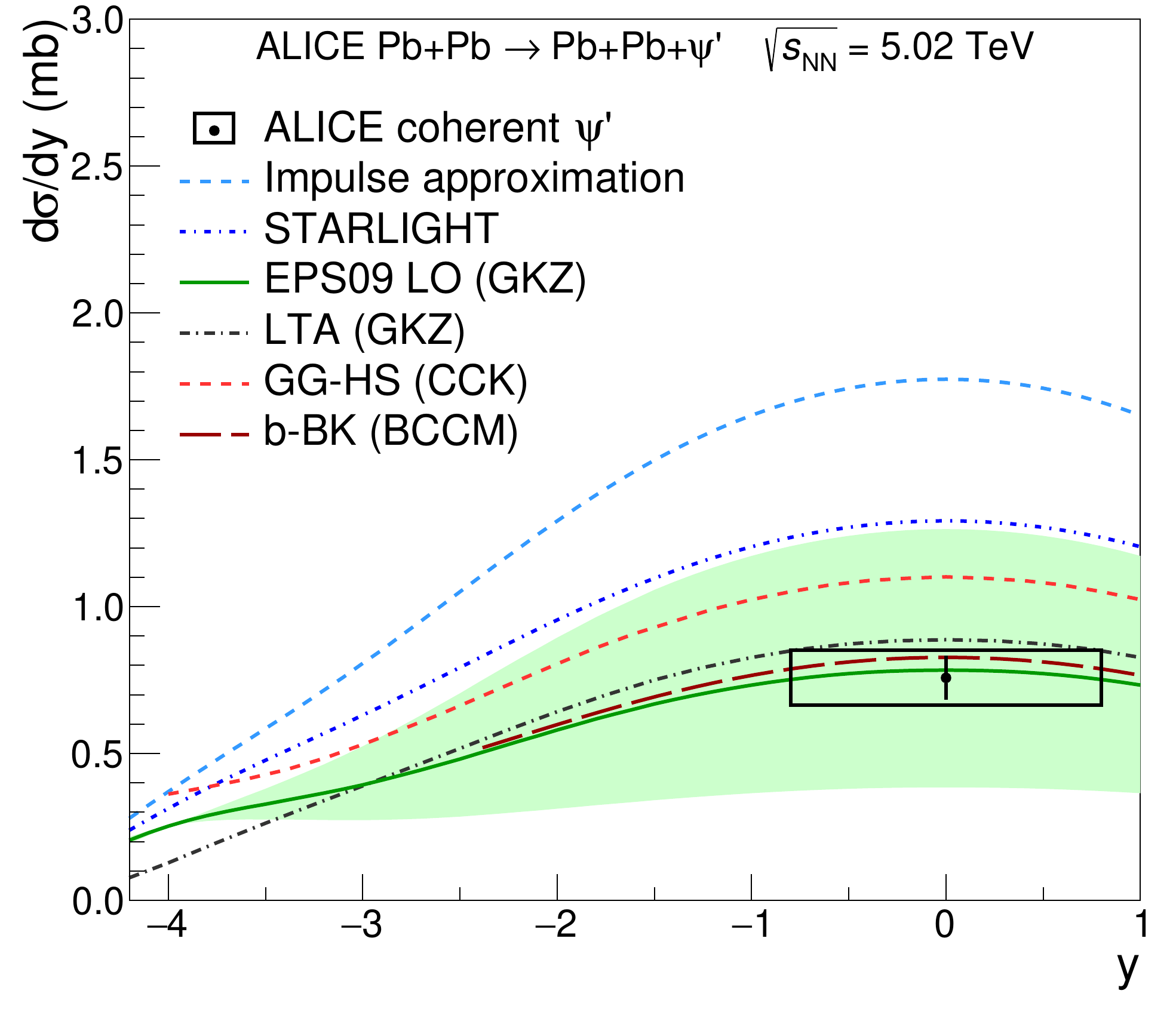}
\caption{Measured differential cross section of the coherent \Jpsi\ (left) and \Ppsi\ (right) photoproduction in \PbPb UPC events. The error bars (boxes) show the statistical (systematic) uncertainties. The theoretical calculations are also shown. The green band represents the uncertainties of the EPS09 LO calculation.}
\label{fig:XsectionPredictions}
\end{figure}

The impulse approximation, taken from STARlight~\cite{PhysRevC.60.014903}, is based on data from exclusive \Jpsi~photoproduction off protons and neglects all nuclear effects except for the coherence. 
The square root of the ratio of experimental cross sections to the impulse approximation is $0.64 \pm 0.04$ for \Jpsi\ and $0.66 \pm 0.06$ for \Ppsi, where statistical and systematic uncertainties of the ALICE measurements  and a conservative 10\% uncertainty on the impulse approximation are added in quadrature. The obtained nuclear suppression factor reflects the magnitude of the nuclear gluon shadowing factor at typical Bjorken-$x$ values in the range $(0.3,1.4)\times10^{-3}$ and is in good agreement with $R_g(x \sim 10^{-3}) = 0.61^{+0.05}_{-0.04}$ obtained  in Ref.~\cite{Guzey:2013xba} from the \Jpsi\ cross section measurement in UPCs at \twosevensixnn.

STARlight is based on the Vector Meson Dominance model and a parametrization of the existing data on \Jpsi\ photoproduction off protons~\cite{Klein:2016yzr}. A Glauber-like formalism is used to calculate the \Jpsi\ photoproduction cross section in \PbPb~UPCs accounting for multiple interactions within the nucleus but not accounting for the gluon shadowing corrections. The STARlight model overpredicts the data indicating that Glauber-like rescatterings alone are not enough to explain the observed suppression of the coherent \Jpsi\ cross section.

Guzey, Kryshen and Zhalov~\cite{Guzey:2016piu} provide two calculations (GKZ), one based on the EPS09 LO parametrization of the available nuclear shadowing data~\cite{Eskola:2009uj} and the other on the leading twist approximation (LTA) of nuclear shadowing based on the combination of the Gribov-Glauber theory and the diffractive PDFs from HERA~\cite{Frankfurt:2011cs}. Both the LTA model and the EPS09 curve, corresponding to the EPS09 LO central set (uncertainties of the EPS09 calculation are represented by the green band), are found to be in a good agreement with the \Jpsi\ and \Ppsi\ cross sections measured at midrapidity. However, these models are in tension with the \Jpsi\ data at semi-forward rapidity in the range $2.5<|y|<3.5$, indicating that the nuclear shadowing might have a smaller effect at Bjorken $x\sim 10^{-2}$  or $x\sim 5\times 10^{-5}$ corresponding to this rapidity range. It is worth noting that the GKZ predictions are based on gluon shadowing effects at a scale $Q^2 = 3\, {\rm GeV^2}$ in contrast to the default value of $2.4 \, {\rm GeV^2}$ which is used in other models and also in LTA predictions at lower energies~\cite{Rebyakova:2011vf}. 
The modified $Q^2$ value was found to provide better description of the coherent \Jpsi\ production cross section in \PbPb UPC measured by ALICE in Run 1 as well as exclusive \Jpsi\ photoproduction off protons~\cite{Guzey:2013qza}.

Calculations by Cepila, Contreras, Krelina and Tapia Takaki (CCK) are based on the colour dipole model with the structure of the nucleon in the transverse plane described by the so-called hot spots, regions of high gluonic density, whose number increases with the increasing energy~\cite{Cepila:2016uku,Cepila:2017nef}.
Nuclear effects are implemented along the ideas proposed in the energy-dependent hot-spot model with the standard Glauber-Gribov formalism (GG-HS) for the extension to the nuclear case. The GG-HS model agrees with the \Jpsi\ measurements at midrapidity and at most forward rapidities but underpredicts them at semi-forward rapidities. The \Ppsi\ measurement at midrapidity is overpredicted by this model.

Calculations by Bendova, Cepila, Contreras, Matas (BCCM) are based on the color dipole approach coupled to the solutions of the impact-parameter dependent Balitsky-Kovchegov equation with initial conditions based on the Woods-Saxon shape of the Pb nucleus~\cite{Bendova:2020hbb}. The model is in a reasonable agreement with the \Jpsi\ and \Ppsi\ data at midrapidity.

Several theory groups provided predictions for \Jpsi\ within the color dipole approach coupled to the Color Glass Condensate (CGC) formalism with different assumptions on the dipole-proton scattering amplitude. Predictions by Gonçalves, Machado et al. (GM)~\cite{PhysRevC.90.015203, Santos:2014zna} based on the IIM and b-CGC models for the scattering amplitude agree with the \Jpsi\ data rather well at midrapidity but strongly underpredict the data at forward rapidities. Predictions by Lappi and Mäntysaari (LM) based on the IPsat model~\cite{PhysRevC.83.065202, PhysRevC.87.032201} overpredict the ALICE measurements at midrapidity, but match them at forward rapidities. Recent predictions by \L{}uszczak and Sch\"afer (LS BGK-I) within the color-dipole formulation of the Glauber-Gribov theory~\cite{Luszczak:2019vdc} are in agreement with the \Jpsi\ data at semi-forward rapidities, $2.5<|y|<3$, slightly underpredict the data at more forward rapidities $3<|y|<4$ and overpredict the data at midrapidity.

The comparison of the current results and those at lower energies of Run 1 may provide interesting insights for model builders and further constraints on gluon shadowing effects. In particular, the ratio of the coherent \Jpsi\ production cross sections at central and forward rapidities at \fivenn :
\begin{equation}
  r(5.02\,{\rm TeV}) = \frac
  {{\rm d}\sigma^{\rm coh}_{J/\psi}/{\rm d}y (|y|<0.8, 5.02\, {\rm TeV})}
  {{\rm d}\sigma^{\rm coh}_{J/\psi}/{\rm d}y (-3.25<y<-3.0, 5.02\, {\rm TeV})}
  = 1.45^{+0.13}_{-0.12}
\end{equation}
is significantly smaller compared to a similar ratio at \twosevensixnn~\cite{Abbas:2013oua}:
\begin{equation}
  r(2.76\,{\rm TeV}) = \frac
  {{\rm d}\sigma^{\rm coh}_{J/\psi}/{\rm d}y (|y|<0.9, 2.76\, {\rm TeV})}
  {{\rm d}\sigma^{\rm coh}_{J/\psi}/{\rm d}y (-3.6<y<-2.6, 2.76\, {\rm TeV})}
  = 2.38^{+0.46}_{-0.38}
\end{equation}
where statistical and systematic uncertainties were added in quadrature. The observed flattening of the shape of the rapidity-differential cross section is qualitatively reproduced in the impulse approximation from STARlight where the central-to-forward ratio decreases from 2.66 at \twosevensixnn to 2.22 at \fivenn mainly due to increased contribution of higher photon-nucleus energies to the measured cross section at forward rapidity. However, the magnitude and the energy dependence of the central-to-forward ratio are significantly modified compared to the impulse approximation due to gluon shadowing effects.

The measured ratio of the \Ppsi\ to \Jpsi\ cross section is compatible 
with the previous ALICE measurement at forward rapidities $R = 0.150 \pm 0.018 \rm(stat.) \pm 0.021 \rm(syst.) \pm 0.007 \rm(BR)$~\cite{Acharya:2019vlb},
with the exclusive photoproduction cross section ratio $R = 0.166 \pm 0.007 \rm(stat.) \pm 0.008 \rm(syst.) \pm 0.007 \rm(BR)$ measured by the H1 collaboration in ep collisions~\cite{Adloff:2002re} and with the ratio $R \approx 0.19$ measured by the LHCb collaboration in pp collisions~\cite{Aaij:2018arx}. The measured ratio also agrees with the ratio $R \approx 0.20$ predicted in the leading twist approximation~\cite{Guzey:2016piu} for \PbPb UPCs at midrapidity. 
The \Ppsi\ to \Jpsi\ coherent cross section ratio is expected to have a mild dependence on the collision energy~\cite{Guzey:2016piu}. Therefore, the measured ratio can be directly compared to the unexpectedly large \Ppsi\ to \Jpsi\ coherent cross section ratio $R = 0.34^{+0.08}_{-0.07}$, measured by ALICE at midrapidity in \PbPb UPCs at \twosevensixnn ~\cite{Adam:2015sia}. The previous measurement is about a factor two larger but is still compatible within 2 standard deviations with the present measurement, owing mainly to the large uncertainties of the previous result.

\section{Conclusions}

The first rapidity-differential measurement on the coherent photoproduction of \Jpsi\ at midrapidity $|y|<0.8$ in \PbPb UPCs at \fivenn has been presented and compared to the model calculations. This data complements the ALICE measurement of the coherent \Jpsi\ cross section at forward rapidity $-4<y<-2.5$ allowing us to provide stringent constraints on nuclear gluon shadowing models.

The nuclear gluon shadowing factor of about 0.64 at Bjorken-$x$ values $x\in(0.3,1.4)\times10^{-3}$ is estimated from the comparison of the measured coherent \Jpsi\ cross section with the impulse approximation at midrapidity. This result is in agreement with the gluon shadowing factor extracted from the previous ALICE measurement of the coherent \Jpsi\ cross section at midrapidity in \PbPb UPCs at \twosevensixnn.

None of the models is able to fully describe the measured forward and central rapidity dependence of the coherent \Jpsi\ cross section. 
The \Jpsi\ measurements at central and most forward rapidities are found to be in agreement with the models based
either on the leading twist approximation of nuclear shadowing, or the central value of the EPS09 parameterization as well as with the energy-dependent hot-spot model extended to the nuclear case by the standard Glauber-Gribov formalism and the color dipole approach coupled to the solutions of the impact-parameter dependent Balitsky-Kovchegov equation. However, these models appear to be in tension with the data at semi-forward rapidities in the range $2.5<|y|<3.5$. The data might be better explained with a model where shadowing has a smaller effect at Bjorken $x\sim 10^{-2}$  or $x\sim 5\times 10^{-5}$ corresponding to this rapidity range.
On the other hand, the models based on the color dipole approach coupled to the color glass condensate formalism describe either the forward or central measurements depending on the dipole scattering amplitude assumptions but they are not able to describe the measured cross section in the full rapidity range. 

The ratio of the \Ppsi\ to \Jpsi\ cross sections at midrapidity is in a reasonable agreement with the ratio of photoproduction cross sections off protons measured by the H1 and LHCb collaborations, with the leading twist approximation predictions for Pb–Pb UPCs as well as with the ALICE measurement at forward rapidities.


\newenvironment{acknowledgement}{\relax}{\relax}
\begin{acknowledgement}
\section*{Acknowledgements}

The ALICE Collaboration would like to thank all its engineers and technicians for their invaluable contributions to the construction of the experiment and the CERN accelerator teams for the outstanding performance of the LHC complex.
The ALICE Collaboration gratefully acknowledges the resources and support provided by all Grid centres and the Worldwide LHC Computing Grid (WLCG) collaboration.
The ALICE Collaboration acknowledges the following funding agencies for their support in building and running the ALICE detector:
A. I. Alikhanyan National Science Laboratory (Yerevan Physics Institute) Foundation (ANSL), State Committee of Science and World Federation of Scientists (WFS), Armenia;
Austrian Academy of Sciences, Austrian Science Fund (FWF): [M 2467-N36] and Nationalstiftung f\"{u}r Forschung, Technologie und Entwicklung, Austria;
Ministry of Communications and High Technologies, National Nuclear Research Center, Azerbaijan;
Conselho Nacional de Desenvolvimento Cient\'{\i}fico e Tecnol\'{o}gico (CNPq), Financiadora de Estudos e Projetos (Finep), Funda\c{c}\~{a}o de Amparo \`{a} Pesquisa do Estado de S\~{a}o Paulo (FAPESP) and Universidade Federal do Rio Grande do Sul (UFRGS), Brazil;
Ministry of Education of China (MOEC) , Ministry of Science \& Technology of China (MSTC) and National Natural Science Foundation of China (NSFC), China;
Ministry of Science and Education and Croatian Science Foundation, Croatia;
Centro de Aplicaciones Tecnol\'{o}gicas y Desarrollo Nuclear (CEADEN), Cubaenerg\'{\i}a, Cuba;
%
Ministry of Education, Youth and Sports of the Czech Republic, Czech Republic; Czech Science Foundation;
The Danish Council for Independent Research | Natural Sciences, the VILLUM FONDEN and Danish National Research Foundation (DNRF), Denmark;
Helsinki Institute of Physics (HIP), Finland;
Commissariat \`{a} l'Energie Atomique (CEA) and Institut National de Physique Nucl\'{e}aire et de Physique des Particules (IN2P3) and Centre National de la Recherche Scientifique (CNRS), France;
Bundesministerium f\"{u}r Bildung und Forschung (BMBF) and GSI Helmholtzzentrum f\"{u}r Schwerionenforschung GmbH, Germany;
General Secretariat for Research and Technology, Ministry of Education, Research and Religions, Greece;
National Research, Development and Innovation Office, Hungary;
Department of Atomic Energy Government of India (DAE), Department of Science and Technology, Government of India (DST), University Grants Commission, Government of India (UGC) and Council of Scientific and Industrial Research (CSIR), India;
Indonesian Institute of Science, Indonesia;
Istituto Nazionale di Fisica Nucleare (INFN), Italy;
Institute for Innovative Science and Technology , Nagasaki Institute of Applied Science (IIST), Japanese Ministry of Education, Culture, Sports, Science and Technology (MEXT) and Japan Society for the Promotion of Science (JSPS) KAKENHI, Japan;
Consejo Nacional de Ciencia (CONACYT) y Tecnolog\'{i}a, through Fondo de Cooperaci\'{o}n Internacional en Ciencia y Tecnolog\'{i}a (FONCICYT) and Direcci\'{o}n General de Asuntos del Personal Academico (DGAPA), Mexico;
Nederlandse Organisatie voor Wetenschappelijk Onderzoek (NWO), Netherlands;
The Research Council of Norway, Norway;
Commission on Science and Technology for Sustainable Development in the South (COMSATS), Pakistan;
Pontificia Universidad Cat\'{o}lica del Per\'{u}, Peru;
Ministry of Science and Higher Education, National Science Centre and WUT ID-UB, Poland;
Korea Institute of Science and Technology Information and National Research Foundation of Korea (NRF), Republic of Korea;
Ministry of Education and Scientific Research, Institute of Atomic Physics and Ministry of Research and Innovation and Institute of Atomic Physics, Romania;
Joint Institute for Nuclear Research (JINR), Ministry of Education and Science of the Russian Federation, National Research Centre Kurchatov Institute, Russian Science Foundation and Russian Foundation for Basic Research, Russia;
Ministry of Education, Science, Research and Sport of the Slovak Republic, Slovakia;
National Research Foundation of South Africa, South Africa;
Swedish Research Council (VR) and Knut \& Alice Wallenberg Foundation (KAW), Sweden;
European Organization for Nuclear Research, Switzerland;
Suranaree University of Technology (SUT), National Science and Technology Development Agency (NSDTA) and Office of the Higher Education Commission under NRU project of Thailand, Thailand;
Turkish Atomic Energy Agency (TAEK), Turkey;
National Academy of  Sciences of Ukraine, Ukraine;
Science and Technology Facilities Council (STFC), United Kingdom;
National Science Foundation of the United States of America (NSF) and United States Department of Energy, Office of Nuclear Physics (DOE NP), United States of America.
\end{acknowledgement}

\bibliographystyle{utphys}   
\bibliography{bibliography}

\providecommand{\href}[2]{#2}\begingroup\raggedright\begin{thebibliography}{10}

\bibitem{Bertulani:2005ru}
C.~A. Bertulani, S.~R. Klein, and J.~Nystrand, ``{Physics of ultra-peripheral
  nuclear collisions}'',
  \href{http://dx.doi.org/10.1146/annurev.nucl.55.090704.151526}{{\em Ann. Rev.
  Nucl. Part. Sci.} {\bfseries 55} (2005) 271--310},
  \href{http://arxiv.org/abs/nucl-ex/0502005}{{\ttfamily
  arXiv:nucl-ex/0502005}}.

\bibitem{Baltz:2007kq}
A.~Baltz, ``{The Physics of Ultraperipheral Collisions at the LHC}'',
  \href{http://dx.doi.org/10.1016/j.physrep.2007.12.001}{{\em Phys. Rept.}
  {\bfseries 458} (2008) 1--171},
  \href{http://arxiv.org/abs/0706.3356}{{\ttfamily arXiv:0706.3356 [nucl-ex]}}.

\bibitem{Contreras:2015dqa}
J.~G. Contreras and J.~D. Tapia~Takaki, ``{Ultra-peripheral heavy-ion
  collisions at the LHC}'',
\href{http://dx.doi.org/10.1142/S0217751X15420129}{{\em Int. J. Mod. Phys.}
  {\bfseries A30} (2015) 1542012}.

\bibitem{Klein:2019qfb}
S.~R. Klein and H.~M\"antysaari, ``{Imaging the nucleus with high-energy
  photons}'', \href{http://dx.doi.org/10.1038/s42254-019-0107-6}{{\em Nature
  Rev. Phys.} {\bfseries 1} no.~11, (2019) 662--674},
  \href{http://arxiv.org/abs/1910.10858}{{\ttfamily arXiv:1910.10858
  [hep-ex]}}.

\bibitem{ALICE:2012aa}
{\bfseries ALICE} Collaboration, B.~Abelev {\em et~al.}, ``{Measurement of the
  Cross Section for Electromagnetic Dissociation with Neutron Emission in Pb-Pb
  Collisions at $\sqrt{s_{NN}}$ = 2.76 TeV}'',
  \href{http://dx.doi.org/10.1103/PhysRevLett.109.252302}{{\em Phys. Rev.
  Lett.} {\bfseries 109} (2012) 252302},
  \href{http://arxiv.org/abs/1203.2436}{{\ttfamily arXiv:1203.2436 [nucl-ex]}}.

\bibitem{Guzey:2018tlk}
V.~Guzey, M.~Strikman, and M.~Zhalov, ``{Nucleon dissociation and incoherent
  \Jpsi\ photoproduction on nuclei in ion ultraperipheral collisions at the
  Large Hadron Collider}'',
  \href{http://dx.doi.org/10.1103/PhysRevC.99.015201}{{\em Phys. Rev. C}
  {\bfseries 99} no.~1, (2019) 015201},
  \href{http://arxiv.org/abs/1808.00740}{{\ttfamily arXiv:1808.00740
  [hep-ph]}}.

\bibitem{Ryskin:1992ui}
M.~Ryskin, ``{Diffractive \Jpsi\ electroproduction in LLA QCD}'',
  \href{http://dx.doi.org/10.1007/BF01555742}{{\em Z. Phys. C} {\bfseries 57}
  (1993) 89--92}.

\bibitem{Armesto:2006ph}
N.~Armesto, ``{Nuclear shadowing}'',
  \href{http://dx.doi.org/10.1088/0954-3899/32/11/R01}{{\em J. Phys. G}
  {\bfseries 32} (2006) R367--R394},
  \href{http://arxiv.org/abs/hep-ph/0604108}{{\ttfamily arXiv:hep-ph/0604108}}.

\bibitem{Frankfurt:2011cs}
L.~Frankfurt, V.~Guzey, and M.~Strikman, ``{Leading Twist Nuclear Shadowing
  Phenomena in Hard Processes with Nuclei}'',
  \href{http://dx.doi.org/10.1016/j.physrep.2011.12.002}{{\em Phys. Rept.}
  {\bfseries 512} (2012) 255--393},
  \href{http://arxiv.org/abs/1106.2091}{{\ttfamily arXiv:1106.2091 [hep-ph]}}.

\bibitem{Bendova:2020hbb}
{D. Bendova and J. Cepila and J.G. Contreras and M. Matas}, ``{Photonuclear
  \Jpsi\ production at the LHC: Proton-based versus nuclear dipole scattering
  amplitudes}'',
  \href{http://dx.doi.org/{https://doi.org/10.1016/j.physletb.2021.136306}}{{\em
  {Physics Letters B}} {\bfseries {817}} ({2021}) {136306}}.
  \url{{https://www.sciencedirect.com/science/article/pii/S037026932100246X}}.

\bibitem{PhysRevC.90.015203}
V.~P. Gon\ifmmode~\mbox{\c{c}}\else \c{c}\fi{}alves, B.~D. Moreira, and F.~S.
  Navarra, ``{Investigation of diffractive photoproduction of \Jpsi\ in
  hadronic collisions}'',
  \href{http://dx.doi.org/10.1103/PhysRevC.90.015203}{{\em Phys. Rev. C}
  {\bfseries 90} (Jul, 2014) 015203}.
  \url{https://link.aps.org/doi/10.1103/PhysRevC.90.015203}.

\bibitem{PhysRevC.83.065202}
T.~Lappi and H.~M\"antysaari, ``{Incoherent diffractive \Jpsi\ production in
  high-energy nuclear deep-inelastic scattering}'',
  \href{http://dx.doi.org/10.1103/PhysRevC.83.065202}{{\em Phys. Rev. C}
  {\bfseries 83} (Jun, 2011) 065202}.
  \url{https://link.aps.org/doi/10.1103/PhysRevC.83.065202}.

\bibitem{Luszczak:2019vdc}
A.~\L{}uszczak and W.~Schäfer, ``{Coherent photoproduction of \Jpsi\ in
  nucleus-nucleus collisions in the color dipole approach}'',
  \href{http://dx.doi.org/10.1103/PhysRevC.99.044905}{{\em Phys. Rev. C}
  {\bfseries 99} no.~4, (2019) 044905},
  \href{http://arxiv.org/abs/1901.07989}{{\ttfamily arXiv:1901.07989
  [hep-ph]}}.

\bibitem{Cepila:2017nef}
J.~Cepila, J.~G. Contreras, and M.~Krelina, ``{Coherent and incoherent \Jpsi\
  photonuclear production in an energy-dependent hot-spot model}'',
  \href{http://dx.doi.org/10.1103/PhysRevC.97.024901}{{\em Phys. Rev. C}
  {\bfseries 97} no.~2, (2018) 024901},
  \href{http://arxiv.org/abs/1711.01855}{{\ttfamily arXiv:1711.01855
  [hep-ph]}}.

\bibitem{Eskola:2009uj}
K.~Eskola, H.~Paukkunen, and C.~Salgado, ``{EPS09: A New Generation of NLO and
  LO Nuclear Parton Distribution Functions}'',
  \href{http://dx.doi.org/10.1088/1126-6708/2009/04/065}{{\em JHEP} {\bfseries
  04} (2009) 065}, \href{http://arxiv.org/abs/0902.4154}{{\ttfamily
  arXiv:0902.4154 [hep-ph]}}.

\bibitem{Eskola:2016oht}
K.~J. Eskola, P.~Paakkinen, H.~Paukkunen, and C.~A. Salgado, ``{EPPS16: Nuclear
  parton distributions with LHC data}'',
  \href{http://dx.doi.org/10.1140/epjc/s10052-017-4725-9}{{\em Eur. Phys. J. C}
  {\bfseries 77} no.~3, (2017) 163},
  \href{http://arxiv.org/abs/1612.05741}{{\ttfamily arXiv:1612.05741
  [hep-ph]}}.

\bibitem{Kovarik:2015cma}
K.~Kovarik {\em et~al.}, ``{nCTEQ15 - Global analysis of nuclear parton
  distributions with uncertainties in the CTEQ framework}'',
  \href{http://dx.doi.org/10.1103/PhysRevD.93.085037}{{\em Phys. Rev. D}
  {\bfseries 93} no.~8, (2016) 085037},
  \href{http://arxiv.org/abs/1509.00792}{{\ttfamily arXiv:1509.00792
  [hep-ph]}}.

\bibitem{AbdulKhalek:2020yuc}
R.~Abdul~Khalek, J.~J. Ethier, J.~Rojo, and G.~van Weelden, ``{nNNPDF2.0: quark
  flavor separation in nuclei from LHC data}'',
  \href{http://dx.doi.org/10.1007/JHEP09(2020)183}{{\em JHEP} {\bfseries 09}
  (2020) 183}, \href{http://arxiv.org/abs/2006.14629}{{\ttfamily
  arXiv:2006.14629 [hep-ph]}}.

\bibitem{Guzey:2013xba}
V.~Guzey, E.~Kryshen, M.~Strikman, and M.~Zhalov, ``{Evidence for nuclear gluon
  shadowing from the ALICE measurements of PbPb ultraperipheral exclusive
  \Jpsi\ production}'',
  \href{http://dx.doi.org/10.1016/j.physletb.2013.08.043}{{\em Phys. Lett. B}
  {\bfseries 726} (2013) 290--295},
  \href{http://arxiv.org/abs/1305.1724}{{\ttfamily arXiv:1305.1724 [hep-ph]}}.

\bibitem{Contreras:2016pkc}
J.~G. Contreras, ``{Gluon shadowing at small $x$ from coherent \Jpsi\
  photoproduction data at energies available at the CERN Large Hadron
  Collider}'', \href{http://dx.doi.org/10.1103/PhysRevC.96.015203}{{\em Phys.
  Rev. C} {\bfseries 96} no.~1, (2017) 015203},
  \href{http://arxiv.org/abs/1610.03350}{{\ttfamily arXiv:1610.03350
  [nucl-ex]}}.

\bibitem{Cepila:2019skb}
J.~Cepila, J.~Nemchik, M.~Krelina, and R.~Pasechnik, ``{Theoretical
  uncertainties in exclusive electroproduction of S-wave heavy quarkonia}'',
  \href{http://dx.doi.org/10.1140/epjc/s10052-019-7016-9}{{\em Eur. Phys. J. C}
  {\bfseries 79} no.~6, (2019) 495},
  \href{http://arxiv.org/abs/1901.02664}{{\ttfamily arXiv:1901.02664
  [hep-ph]}}.

\bibitem{Krelina:2020bxt}
M.~Krelina and J.~Nemchik, ``{D-wave effects in heavy quarkonium production in
  ultraperipheral nuclear collisions}'',
  \href{http://dx.doi.org/10.1103/PhysRevD.102.114033}{{\em Phys. Rev. D}
  {\bfseries 102} no.~11, (2020) 114033},
  \href{http://arxiv.org/abs/2010.00329}{{\ttfamily arXiv:2010.00329
  [hep-ph]}}.

\bibitem{Abelev:2012ba}
{\bfseries ALICE} Collaboration, B.~Abelev {\em et~al.}, ``{Coherent \Jpsi\
  photoproduction in ultra-peripheral Pb-Pb collisions at $\sqrt{s_{NN}} =
  2.76$ TeV}'', \href{http://dx.doi.org/10.1016/j.physletb.2012.11.059}{{\em
  Phys. Lett. B} {\bfseries 718} (2013) 1273--1283},
  \href{http://arxiv.org/abs/1209.3715}{{\ttfamily arXiv:1209.3715 [nucl-ex]}}.

\bibitem{Abbas:2013oua}
{\bfseries ALICE} Collaboration, E.~Abbas {\em et~al.}, ``{Charmonium and
  $e^+e^-$ pair photoproduction at mid-rapidity in ultra-peripheral Pb-Pb
  collisions at $\sqrt{s_{\rm NN}}$=2.76 TeV}'',
  \href{http://dx.doi.org/10.1140/epjc/s10052-013-2617-1}{{\em Eur. Phys. J. C}
  {\bfseries 73} no.~11, (2013) 2617},
  \href{http://arxiv.org/abs/1305.1467}{{\ttfamily arXiv:1305.1467 [nucl-ex]}}.

\bibitem{Adam:2015sia}
{\bfseries ALICE} Collaboration, J.~Adam {\em et~al.}, ``{Coherent $\psi$(2S)
  photo-production in ultra-peripheral Pb--Pb collisions at $\sqrt{s}_{\rm NN}$
  = 2.76 TeV}'', \href{http://dx.doi.org/10.1016/j.physletb.2015.10.040}{{\em
  Phys. Lett. B} {\bfseries 751} (2015) 358--370},
  \href{http://arxiv.org/abs/1508.05076}{{\ttfamily arXiv:1508.05076
  [nucl-ex]}}.

\bibitem{Khachatryan:2016qhq}
{\bfseries CMS} Collaboration, V.~Khachatryan {\em et~al.}, ``{Coherent \Jpsi\
  photoproduction in ultra-peripheral PbPb collisions at $\sqrt {s_{NN}} =$
  2.76 TeV with the CMS experiment}'',
  \href{http://dx.doi.org/10.1016/j.physletb.2017.07.001}{{\em Phys. Lett. B}
  {\bfseries 772} (2017) 489--511},
  \href{http://arxiv.org/abs/1605.06966}{{\ttfamily arXiv:1605.06966
  [nucl-ex]}}.

\bibitem{Acharya:2019vlb}
{\bfseries ALICE} Collaboration, S.~Acharya {\em et~al.}, ``{Coherent \Jpsi\
  photoproduction at forward rapidity in ultra-peripheral Pb-Pb collisions at
  $\sqrt{s_{\rm{NN}}}=5.02$ TeV}'',
  \href{http://dx.doi.org/10.1016/j.physletb.2019.134926}{{\em Phys. Lett.}
  {\bfseries B798} (2019) 134926},
\href{http://arxiv.org/abs/1904.06272}{{\ttfamily arXiv:1904.06272 [nucl-ex]}}.

\bibitem{Aamodt:2008zz}
{\bfseries ALICE} Collaboration, K.~Aamodt {\em et~al.}, ``{The ALICE
  experiment at the CERN LHC}'',
  \href{http://dx.doi.org/10.1088/1748-0221/3/08/S08002}{{\em JINST} {\bfseries
  3} (2008) S08002}.

\bibitem{Abelev:2014ffa}
{\bfseries ALICE} Collaboration, B.~B. Abelev {\em et~al.}, ``{Performance of
  the ALICE Experiment at the CERN LHC}'',
  \href{http://dx.doi.org/10.1142/S0217751X14300440}{{\em Int. J. Mod. Phys. A}
  {\bfseries 29} (2014) 1430044},
  \href{http://arxiv.org/abs/1402.4476}{{\ttfamily arXiv:1402.4476 [nucl-ex]}}.

\bibitem{Aamodt:2010aa}
{\bfseries ALICE} Collaboration, K.~Aamodt {\em et~al.}, ``{Alignment of the
  ALICE Inner Tracking System with cosmic-ray tracks}'',
  \href{http://dx.doi.org/10.1088/1748-0221/5/03/P03003}{{\em JINST} {\bfseries
  5} (2010) P03003}, \href{http://arxiv.org/abs/1001.0502}{{\ttfamily
  arXiv:1001.0502 [physics.ins-det]}}.

\bibitem{Alme_2010}
J.~Alme, Y.~Andres, H.~Appelshäuser, S.~Bablok, N.~Bialas, R.~Bolgen,
  U.~Bonnes, R.~Bramm, P.~Braun-Munzinger, R.~Campagnolo, and et~al., ``{The
  ALICE TPC, a large 3-dimensional tracking device with fast readout for
  ultra-high multiplicity events}'',
  \href{http://dx.doi.org/10.1016/j.nima.2010.04.042}{{\em Nuclear Instruments
  and Methods in Physics Research Section A: Accelerators, Spectrometers,
  Detectors and Associated Equipment} {\bfseries 622} no.~1, (Oct, 2010)
  316–367}. \url{http://dx.doi.org/10.1016/j.nima.2010.04.042}.

\bibitem{Akindinov:2013tea}
A.~Akindinov {\em et~al.}, ``{Performance of the ALICE Time-Of-Flight detector
  at the LHC}'', \href{http://dx.doi.org/10.1140/epjp/i2013-13044-x}{{\em Eur.
  Phys. J. Plus} {\bfseries 128} (2013) 44}.

\bibitem{Abbas:2013taa}
{\bfseries ALICE} Collaboration, E.~Abbas {\em et~al.}, ``{Performance of the
  ALICE VZERO system}'',
  \href{http://dx.doi.org/10.1088/1748-0221/8/10/P10016}{{\em JINST} {\bfseries
  8} (2013) P10016}, \href{http://arxiv.org/abs/1306.3130}{{\ttfamily
  arXiv:1306.3130 [nucl-ex]}}.

\bibitem{Akiba:2016ofq}
{\bfseries LHC Forward Physics Working Group} Collaboration, K.~Akiba {\em
  et~al.}, ``{LHC Forward Physics}'',
  \href{http://dx.doi.org/10.1088/0954-3899/43/11/110201}{{\em J. Phys. G}
  {\bfseries 43} (2016) 110201},
  \href{http://arxiv.org/abs/1611.05079}{{\ttfamily arXiv:1611.05079
  [hep-ph]}}.

\bibitem{ALICE-PUBLIC-2021-001}
{\bfseries ALICE} Collaboration, S.~Acharya {\em et~al.}, ``{ALICE luminosity
  determination for Pb–Pb collisions at $\sqrt {s_{\rm NN}} =$ 5.02 TeV}'',
  tech. rep., CERN, January, 2021.
\newblock \url{https://cds.cern.ch/record/0000}.

\bibitem{Pshenichnov:2011zz}
I.~A. Pshenichnov, ``{Electromagnetic excitation and fragmentation of
  ultrarelativistic nuclei}'',
\href{http://dx.doi.org/10.1134/S1063779611020067}{{\em Phys. Part. Nucl.}
  {\bfseries 42} (2011) 215--250}.

\bibitem{Klein:2016yzr}
S.~R. Klein, J.~Nystrand, J.~Seger, Y.~Gorbunov, and J.~Butterworth,
  ``{STARlight: A Monte Carlo simulation program for ultra-peripheral
  collisions of relativistic ions}'',
  \href{http://dx.doi.org/10.1016/j.cpc.2016.10.016}{{\em Comput. Phys.
  Commun.} {\bfseries 212} (2017) 258--268},
  \href{http://arxiv.org/abs/1607.03838}{{\ttfamily arXiv:1607.03838
  [hep-ph]}}.

\bibitem{Brun:1082634}
R.~Brun, F.~Bruyant, F.~Carminati, S.~Giani, M.~Maire, A.~McPherson,
  G.~Patrick, and L.~Urban,
  \href{http://dx.doi.org/10.17181/CERN.MUHF.DMJ1}{{\em {GEANT: Detector
  Description and Simulation Tool; Oct 1994}}}.
\newblock CERN Program Library. CERN, Geneva, 1993.
\newblock \url{http://cds.cern.ch/record/1082634}.
\newblock Long Writeup W5013.

\bibitem{Alexa:2013xxa}
{\bfseries H1} Collaboration, C.~Alexa {\em et~al.}, ``{Elastic and
  Proton-Dissociative Photoproduction of \Jpsi\ Mesons at HERA}'',
  \href{http://dx.doi.org/10.1140/epjc/s10052-013-2466-y}{{\em Eur. Phys. J.}
  {\bfseries C73} no.~6, (2013) 2466},
\href{http://arxiv.org/abs/1304.5162}{{\ttfamily arXiv:1304.5162 [hep-ex]}}.

\bibitem{Adloff:2002re}
{\bfseries H1} Collaboration, C.~Adloff {\em et~al.}, ``{Diffractive
  photoproduction of $\psi$(2S) mesons at HERA}'',
  \href{http://dx.doi.org/10.1016/S0370-2693(02)02275-X}{{\em Phys. Lett. B}
  {\bfseries 541} (2002) 251--264},
  \href{http://arxiv.org/abs/hep-ex/0205107}{{\ttfamily arXiv:hep-ex/0205107}}.

\bibitem{Chekanov:2002xi}
{\bfseries ZEUS} Collaboration, S.~Chekanov {\em et~al.}, ``{Exclusive
  photoproduction of J / psi mesons at HERA}'',
  \href{http://dx.doi.org/10.1007/s10052-002-0953-7}{{\em Eur. Phys. J. C}
  {\bfseries 24} (2002) 345--360},
  \href{http://arxiv.org/abs/hep-ex/0201043}{{\ttfamily arXiv:hep-ex/0201043}}.

\bibitem{PhysRevD.62.032002}
{\bfseries BES} Collaboration, J.~Z. Bai {\em et~al.}, ``{$\psi({\rm 2S})
  \rightarrow \pi^+\pi^- {\mathrm J}/\psi$ decay distributions}'',
  \href{http://dx.doi.org/10.1103/PhysRevD.62.032002}{{\em Phys. Rev. D}
  {\bfseries 62} (Jul, 2000) 032002}.
  \url{https://link.aps.org/doi/10.1103/PhysRevD.62.032002}.

\bibitem{Ducati:2013bya}
M.~B.~G. Ducati, M.~Griep, and M.~Machado, ``{Diffractive photoproduction of
  radially excited psi(2S) mesons in photon-Pomeron reactions in PbPb
  collisions at the CERN LHC}'',
  \href{http://dx.doi.org/10.1103/PhysRevC.88.014910}{{\em Phys. Rev. C}
  {\bfseries 88} (2013) 014910},
  \href{http://arxiv.org/abs/1305.2407}{{\ttfamily arXiv:1305.2407 [hep-ph]}}.

\bibitem{Zyla:2020zbs}
{\bfseries Particle Data Group} Collaboration, P.~Zyla {\em et~al.}, ``{Review
  of Particle Physics}'', \href{http://dx.doi.org/10.1093/ptep/ptaa104}{{\em
  PTEP} {\bfseries 2020} no.~8, (2020) 083C01}.

\bibitem{PhysRevC.60.014903}
S.~R. Klein and J.~Nystrand, ``Exclusive vector meson production in
  relativistic heavy ion collisions'',
  \href{http://dx.doi.org/10.1103/PhysRevC.60.014903}{{\em Phys. Rev. C}
  {\bfseries 60} (1999) 014903}.
  \url{https://link.aps.org/doi/10.1103/PhysRevC.60.014903}.

\bibitem{Guzey:2016piu}
V.~Guzey, E.~Kryshen, and M.~Zhalov, ``{Coherent photoproduction of vector
  mesons in ultraperipheral heavy ion collisions: Update for run 2 at the CERN
  Large Hadron Collider}'',
  \href{http://dx.doi.org/10.1103/PhysRevC.93.055206}{{\em Phys. Rev. C}
  {\bfseries 93} no.~5, (2016) 055206},
  \href{http://arxiv.org/abs/1602.01456}{{\ttfamily arXiv:1602.01456
  [nucl-th]}}.

\bibitem{Rebyakova:2011vf}
V.~Rebyakova, M.~Strikman, and M.~Zhalov, ``{Coherent $\rho$ and \Jpsi\
  photoproduction in ultraperipheral processes with electromagnetic
  dissociation of heavy ions at RHIC and LHC}'',
  \href{http://dx.doi.org/10.1016/j.physletb.2012.03.041}{{\em Phys. Lett. B}
  {\bfseries 710} (2012) 647--653},
  \href{http://arxiv.org/abs/1109.0737}{{\ttfamily arXiv:1109.0737 [hep-ph]}}.

\bibitem{Guzey:2013qza}
V.~Guzey and M.~Zhalov, ``{Exclusive $J/{\psi}$ production in ultraperipheral
  collisions at the LHC: constrains on the gluon distributions in the proton
  and nuclei}'', \href{http://dx.doi.org/10.1007/JHEP10(2013)207}{{\em JHEP}
  {\bfseries 10} (2013) 207}, \href{http://arxiv.org/abs/1307.4526}{{\ttfamily
  arXiv:1307.4526 [hep-ph]}}.

\bibitem{Cepila:2016uku}
J.~Cepila, J.~G. Contreras, and J.~D. Tapia~Takaki, ``{Energy dependence of
  dissociative \Jpsi\ photoproduction as a signature of gluon saturation at the
  LHC}'', \href{http://dx.doi.org/10.1016/j.physletb.2016.12.063}{{\em Phys.
  Lett. B} {\bfseries 766} (2017) 186--191},
  \href{http://arxiv.org/abs/1608.07559}{{\ttfamily arXiv:1608.07559
  [hep-ph]}}.

\bibitem{Santos:2014zna}
G.~Sampaio~dos Santos and M.~Machado, ``{On theoretical uncertainty of color
  dipole phenomenology in the \Jpsi\ and $\Upsilon$ photoproduction in pA and
  AA collisions at the CERN Large Hadron Collider}'',
  \href{http://dx.doi.org/10.1088/0954-3899/42/10/105001}{{\em J. Phys. G}
  {\bfseries 42} no.~10, (2015) 105001},
  \href{http://arxiv.org/abs/1411.7918}{{\ttfamily arXiv:1411.7918 [hep-ph]}}.

\bibitem{PhysRevC.87.032201}
T.~Lappi and H.~M\"antysaari, ``{\Jpsi\ production in ultraperipheral Pb+Pb and
  $p$+Pb collisions at energies available at the CERN Large Hadron Collider}'',
  \href{http://dx.doi.org/10.1103/PhysRevC.87.032201}{{\em Phys. Rev. C}
  {\bfseries 87} (Mar, 2013) 032201}.
  \url{https://link.aps.org/doi/10.1103/PhysRevC.87.032201}.

\bibitem{Aaij:2018arx}
{\bfseries LHCb} Collaboration, R.~Aaij {\em et~al.}, ``{Central exclusive
  production of \Jpsi\ and $\psi(2S)$ mesons in $pp$ collisions at
  $\sqrt{s}=13~$TeV}'', \href{http://dx.doi.org/10.1007/JHEP10(2018)167}{{\em
  JHEP} {\bfseries 10} (2018) 167},
  \href{http://arxiv.org/abs/1806.04079}{{\ttfamily arXiv:1806.04079
  [hep-ex]}}.

\end{thebibliography}\endgroup

\newpage
\appendix

%
%

\section{The ALICE Collaboration}
\label{app:collab}

\begin{flushleft} 

S.~Acharya$^{\rm 142}$, 
D.~Adamov\'{a}$^{\rm 97}$, 
A.~Adler$^{\rm 75}$, 
J.~Adolfsson$^{\rm 82}$, 
G.~Aglieri Rinella$^{\rm 35}$, 
M.~Agnello$^{\rm 31}$, 
N.~Agrawal$^{\rm 55}$, 
Z.~Ahammed$^{\rm 142}$, 
S.~Ahmad$^{\rm 16}$, 
S.U.~Ahn$^{\rm 77}$
% ...

\bigskip

\bigskip 

\textbf{\Large Affiliation Notes}

\bigskip 

% ...
$^{\rm V}$ Also at: Institute of Theoretical Physics, University of Wroclaw, Poland\\

\bigskip

\bigskip 

\textbf{\Large Collaboration Institutes}

\bigskip 

% ...
$^{16}$ Department of Physics, Aligarh Muslim University, Aligarh, India\\
% ...
$^{31}$ Dipartimento DISAT del Politecnico and Sezione INFN, Turin, Italy\\
% ...
$^{35}$ European Organization for Nuclear Research (CERN), Geneva, Switzerland\\
% ...
$^{55}$ INFN, Sezione di Bologna, Bologna, Italy\\
% ...
$^{75}$ Johann-Wolfgang-Goethe Universit\"{a}t Frankfurt Institut f\"{u}r Informatik, Fachbereich Informatik und Mathematik, Frankfurt, Germany\\
% ...
$^{77}$ Korea Institute of Science and Technology Information, Daejeon, Republic of Korea\\
% ...
$^{82}$ Lund University Department of Physics, Division of Particle Physics, Lund, Sweden\\
% ...
$^{97}$ Nuclear Physics Institute of the Czech Academy of Sciences, \v{R}e\v{z} u Prahy, Czech Republic\\
% ...
$^{142}$ Variable Energy Cyclotron Centre, Homi Bhabha National Institute, Kolkata, India\\
% ...

\end{flushleft} 
  
\end{document}